\documentclass[aip,pop, reprint, showpacs,superscriptaddress]{revtex4-1}

\usepackage{amsmath,amsfonts,amssymb,amsbsy,amscd}
\usepackage{ifthen}
\usepackage[usenames,dvipsnames]{color}
\usepackage{hyperref}
\usepackage{graphicx}
\usepackage[normalem]{ulem}

\hypersetup{
   pdfauthor=Siminos,
   pdftitle=Kinetic effects on the transition to relativistic self-induced transparency in laser-driven ion acceleration
}

\newcommand{\rf}     [1] {~\cite{#1}}
\newcommand{\refref} [1] {Ref.~\cite{#1}}

\newcommand{\reffig} [1] {Fig.~\ref{#1}}

\newcommand{\refFig} [1] {Figure~\ref{#1}}

\newcommand{\refeq}  [1] {Eq.~(\ref{#1})}
\newcommand{\refEq}  [1] {Eq.~(\ref{#1})}
\newcommand{\refsect}[1]{Sec.~\ref{#1}}
\newcommand{\ie}{{i.e.}}

\newcommand{\nceff}{\ensuremath{n_c^{\mathrm{eff}}}} % Kaw and Dawson threshold density. Was \ensuremath{n_{\mathrm{th}}}
\newcommand{\tl}{\ensuremath{\tau_L}} % optical cycle
\newcommand{\tr}{\ensuremath{\tau_r}} % rising time
\newcommand{\pcr}{\ensuremath{p_x^{\mathrm{cr}}}} % Critical p_x of separatrix
\newcommand{\rsit}{RSIT}
\newcommand{\hb}{HB}
\newcommand{\transient}{dynamic transition}

\newcommand{\vf}{\ensuremath{v_f}}
\newcommand{\xf}{\ensuremath{x_f}}

\newcommand{\vsit}{\ensuremath{v_{\mathrm{SIT}}}}
\newcommand{\vhb}{\ensuremath{v_{\mathrm{HB}}}}
\newcommand{\Ehb}{\ensuremath{\mathcal{E}_{\mathrm{HB}}}}

\newcommand{\Ssit}{\ensuremath{S}} % Cross-correlation function %\ensuremath{S_{\mathrm{SIT}}}

\newcommand{\nicetilde}{\raise.17ex\hbox{$\scriptstyle\sim$}}

\newcommand{\g}[1]{\mbox{\boldmath $#1$}}

\definecolor{hreflinkcolor}{rgb}{0.13,0.17,0.83}
\hypersetup{colorlinks=true,urlcolor=hreflinkcolor,
		linkcolor=hreflinkcolor,citecolor=hreflinkcolor}

\usepackage[normalem]{ulem}

\newcommand{\chalmers}{Department of Physics, Chalmers University of Technology, Gothenburg, Sweden}

\begin{document}

\author{E.~Siminos}
\email{siminos@chalmers.se}
\affiliation{\chalmers}
\author{M.~Grech}\affiliation{LULI, CNRS, UPMC, Ecole Polytechnique, CEA, 91128 Palaiseau, France}
\author{B.~Svedung Wettervik}\affiliation{\chalmers}
\author{T.~F\"ul\"op}\affiliation{\chalmers}

\title{Kinetic and finite ion mass effects on the transition to relativistic self-induced transparency in laser-driven ion acceleration}

\date{\today}

%%%%%%%%%%%%%%%%%%%%%%%%%%%%%%%%%%%%%%%%%%%%%%%%%%%%%%%%%%%%%%%%%%%%%%%%%%%%%%%%
\begin{abstract}
  We study kinetic effects responsible for the transition to relativistic self-induced transparency in the interaction of a circularly-polarized laser-pulse with an overdense plasma and their relation to hole-boring and ion acceleration. {It is demonstrated using particle-in-cell simulations and an analysis of separatrices in single-electron phase-space, that ion motion can suppress fast electron escape to the vacuum, which would otherwise lead to  transition to the relativistic transparency regime. A simple analytical estimate shows that for large laser pulse amplitude $a_0$ the time scale over which ion motion becomes important is much shorter than usually anticipated. As a result, the threshold density above which hole-boring occurs decreases with the charge-to-mass ratio.} Moreover, the transition threshold is seen to depend on the laser temporal profile, due to the effect that the latter has on electron heating. Finally, we report a new regime in which a transition from relativistic transparency   to hole-boring occurs dynamically during the course of the interaction. It is shown that, for a fixed laser intensity, this dynamic transition regime allows optimal ion acceleration in terms of both energy and energy spread.
\end{abstract}
%%%%%%%%%%%%%%%%%%%%%%%%%%%%%%%%%%%%%%%%%%%%%%%%%%%%%%%%%%%%%%%%%%%%%%%%%%%%%%%%

% 52.20.Dq 	Particle orbits 
% 52.35.Mw 	Nonlinear phenomena: waves, wave propagation, and other interactions 
% 		(including parametric effects, mode coupling, ponderomotive effects, etc.) 
% 52.38.-r 	Laser-plasma interactions 
% or
% 52.38.Dx 	Laser light absorption in plasmas (collisional, parametric, etc.)
% 42.25.Gy 	Edge and boundary effects; reflection and refraction 

\maketitle

\section{Introduction}
  
Modern high intensity laser technology 
has made the regime of \emph{relativistic optics} 
experimentally accessible.
In this regime electrons interacting with the laser-field
gain relativistic velocities within an optical cycle
and their motion becomes highly non-linear. 
Exploiting complex laser-plasma interaction
in this regime has led to a wealth of novel applications
ranging from charged particle acceleration~\cite{mourou2006,esarey2009,macchi}
to sources of ultra-short radiation~\cite{teubner2009,corde2013}.

It has long been recognized that in the relativistic optics regime 
even the most basic properties of a plasma such as its
index of refraction are profoundly affected by nonlinearities
in electron motion~\cite{akhiezer1956, kaw1970}.
In particular, the increase of the effective electron mass due to its $\gamma$-factor dependence on the laser
normalized vector potential $a_0=eA_0/(m_ec)$ leads to an effective increase
of the critical density\footnote{By our choice of 
normalization of the incident laser pulse vector potential, \refeq{eq:incident},
this form for $\nceff$ is valid for both circular and linear polarization, if the cycle-averaged $\gamma$-factor is used for the latter.}
\begin{equation}\label{eq:nceff}
  \nceff=\sqrt{1+\frac{a_0^2}{2}}\,n_c\,. 
\end{equation}
{Here $n_c=\epsilon_0{}m_e\omega_L^2/e^2$ is the classical critical
density above which a plasma is nominally opaque to a laser pulse with angular frequency $\omega_L$,
$m_e$ and $-e$ are the electron mass and charge, respectively, 
$c$ is the speed of light in vacuum, and $\epsilon_0$ is the permittivity of free space.}

This simple form for the relativistic critical density
$\nceff$ holds for plane waves {propagating through a uniform and} infinitely long plasma {independently
of their polarization provided that the laser wave amplitude $a_0$ relates to the wave intensity as 
$I_L \lambda_L^2 \simeq  1.38\,a_0^2 \times 10^{18} {\rm W/cm^2\,\mu m^2}$,
with $\lambda_L = 2\pi c/\omega_L$ the laser wavelength.
This effective increase of the critical density} is the basis of the effect known as relativistic self-induced
transparency (\rsit), in which a relativistically intense laser pulse ($a_0\gtrsim1$)
can propagate in a nominally overdense plasma.

\begin{figure}[ht!]
 \begin{center}
  \includegraphics[width=0.8\columnwidth]{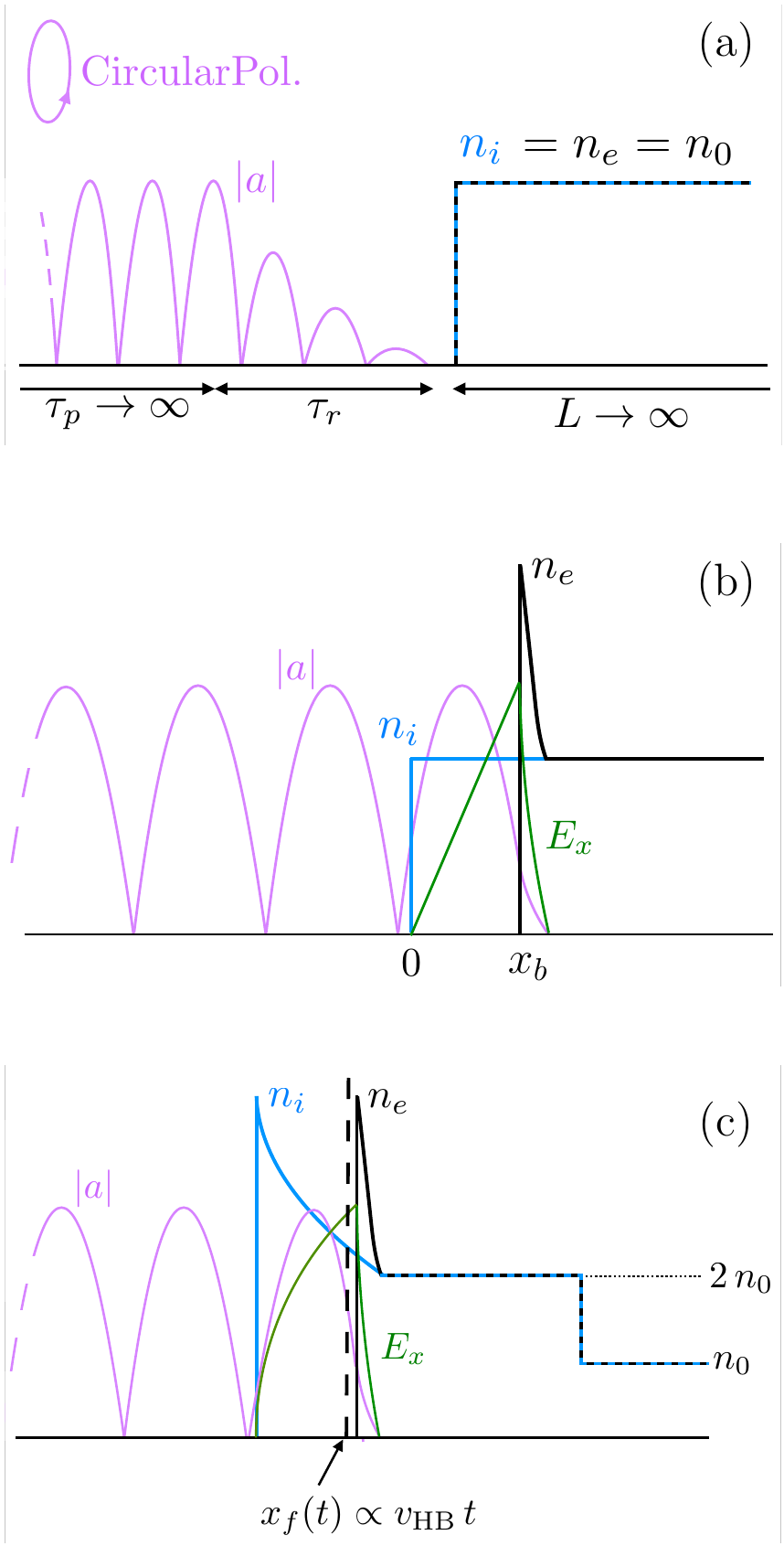}
  \caption{\label{f:intro}
  {(a) Schematic representation of the interaction setup for a hydrogen plasma, showing 
  the electric field $E_x(x)$, vector potential amplitude $|a(x)|$ 
  and ion and electron densities $n_i$ and $n_e$, respectively. 
  (b) Ignoring ion motion a standing wave solution is predicted by cold fluid theory for $n_0>n_{\mathrm{SW}}$, given by \refeq{eq:nCappr}.
  (c) Schematic representation of the \hb\ configuration.}
  }
 \end{center}
\end{figure}

However, when one considers {a laser pulse incident on a bounded plasma,
the situation is much more complicated. In order to allow insight into the basic physical mechanisms 
involved and to establish connection with previous works we consider a simplified 1-dimensional geometry.
We consider a circularly polarized (CP) laser pulse with finite rise time $\tau_r$
and semi-infinite duration, normally incident onto a semi-infinite 
plasma with a constant electron density $n_0 > n_c$, and a sharp interface
with vacuum, see~\reffig{f:intro}(a). This configuration is of particular interest for ultra-high 
contrast laser interaction with thick targets. Since no pre-plasma is assumed, the incoming 
laser pulse interacts directly with a nominally overdense plasma.}  
The ponderomotive force pushes electrons
deeper into the plasma, creating a high-density peak {(compressed electron layer)} 
that may prevent the pulse from propagating further, 
\reffig{f:intro}(b). 
For linearly polarized pulses the strong $\mathbf{J\times B}$ electron heating
can lead to the destruction of the electron density peak
and, to a good approximation, the threshold for RSIT 
is found to be in agreement with $\nceff$~\cite{lefebvre1995,palaniyappan2012}.
{By contrast, for CP pulses, the ponderomotive force is quasi-steady
and electron heating is reduced. As a result, the compressed electron layer forms, 
efficiently reflecting the incident laser pulse. An equilibrium between the ponderomotive and
charge-separation forces {is reached and a standing wave is formed, with the plasma boundary displaced at {a new (time-independent) position} $x_b$, \reffig{f:intro}(b)~\cite{marburger1975}.} 
This situation can be described in the framework of (stationary) cold-fluid theory~\rf{cattani2000,goloviznin2000},
and the existence of a standing wave solution defines the opaque regime of interaction.
It is found that} {a plasma of a given density $n_0$ is opaque (self-shutters) for $a_0$ smaller than a 
threshold amplitude $a_\mathrm{SW}(n_0)$ such that:
\begin{equation}\label{eq:cattani}
   a_\mathrm{SW}^2 = \overline{n}_0\,(1+a_B^2)\,\left(\sqrt{1+a_B^2} -1 \right) - a_B^4/2\,,
\end{equation}
where 
\begin{equation}
  a_B^2 = \overline{n}_0\,\left( \frac{9}{8}\, \overline{n}_0 - 1 + \frac{3}{2}\, \sqrt{\frac{9}{16}\, \overline{n}_0^2 - \overline{n}_0 + 1} \right)\,,
\end{equation}
and $\overline{n}_0\equiv n_0/n_c$.
In the limit of high densities $n_0\gg n_c$ we can invert these expressions to obtain\rf{siminos2012}
the density threshold for the existence of a standing wave
\begin{equation}\label{eq:nCappr}
  n_{\mathrm{SW}}(a_0) \simeq\frac{2}{9} \left(3 + \sqrt{9 \sqrt{6}\, a_0 - 12}\,\right)\,n_c\,.
\end{equation}
Equation~\eqref{eq:nCappr} is plotted in \reffig{f:intro}, and,  for the range $a_0=5-25$ considered here, it is in excellent
agreement with the exact expression~\refeq{eq:cattani}.
{Conversely, cold-fluid theory~\cite{cattani2000,goloviznin2000,eremin2010} predicts that \rsit\ occurs for $n_0<n_{\mathrm{SW}}$.
Note that \refEq{eq:nCappr}} implies a different $a_0\gg1$ scaling for the transition to \rsit, $n_{\mathrm{SW}}\propto{}a_0^{1/2}$, than
\refeq{eq:nceff} which gives $n_c^{\rm eff}\propto a_0$.
}

\begin{figure}[hb!]
%   \begin{center}
  \includegraphics[width=0.8\columnwidth]{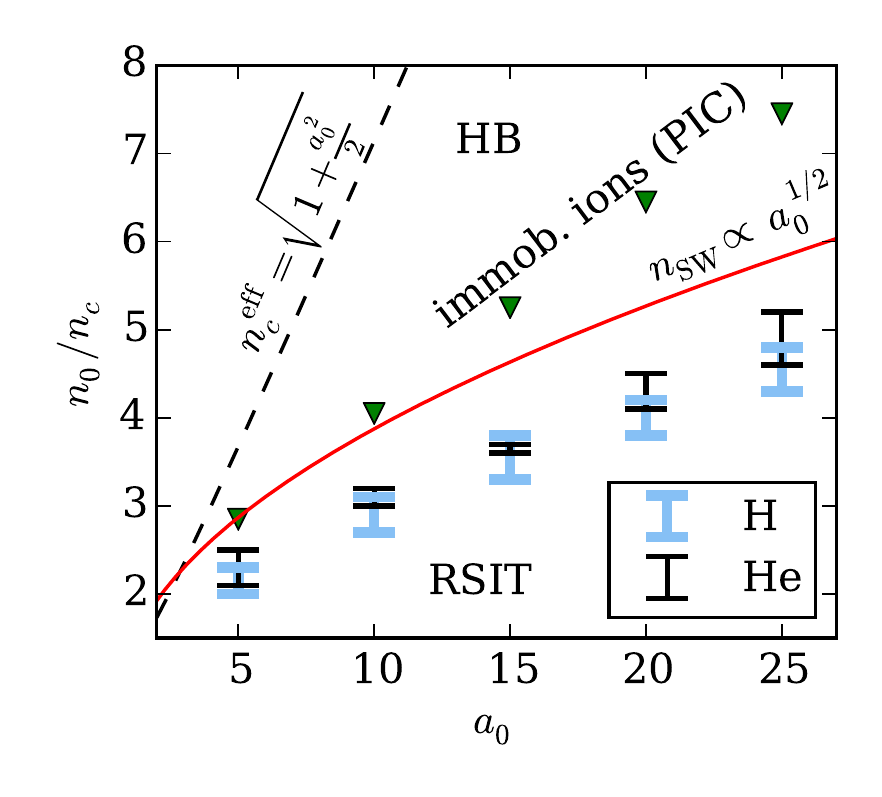}
  \caption{\label{f:fluidPIC}
  Different transition thresholds for \rsit. 
  The black dashed line indicates the transition boundary for infinite plane waves $\nceff=\sqrt{1+a_0^2/2}$.
  The red solid line is the cold-fluid threshold for existence of a standing wave, $n_\mathrm{SW}$, given by \refeq{eq:nCappr}.
  The green triangles indicate the results for the transition threshold $n_{\mathrm{th}}$ 	from PIC simulations with 
  immobile ions, see \refsect{s:detection}. The lower end of the error bars indicates the 	boundary of the \rsit\
  regime for hydrogen and helium, as determined by PIC simulations (\refsect{s:detection}). 
  The upper end of the error bars indicates the boundary of the \hb\ regime.
  The dynamic transition regime lies within the width 
  of the error bars.
  For all PIC simulations
  in this plot a laser pulse rise time $\tr=4\tl$ was used.}
%   \end{center}
\end{figure}

Nevertheless, PIC simulations have shown that 
even modest  electron heating during the early stages of the interaction
can disturb the plasma vacuum interface leading
to a linear scaling for the density transition threshold, $n_{\mathrm{th}}\propto{}a_0$ with a coefficient that depends on the details
of the interaction~\cite{siminos2012}, see {triangles in} \reffig{f:fluidPIC}.

{In addition, ion motion (finite ion mass) has also been found} to lower $n_{\mathrm{th}}$ significantly in PIC simulations\rf{weng2012}.
However, the exact mechanism {responsible for this reduction} has not yet been clarified. 
Determining the conditions and mechanisms responsible 
for transition from the opaque to the \rsit\ regime using CP light is of paramount importance
as it determines the efficiency of laser energy coupling to the plasma,
while it is also crucial for a wide range of applications. {For example, relativistic transparency can be exploited
to enhance the characteristics of laser-pulses~\cite{bin2015}, it may affect the propagation of probe pulses 
in plasmas with fast particles~\cite{stark2015,saevert2015,siminos2016} 
and {has led to the development of novel ion acceleration schemes~\cite{Yin_LPB_2006,willingale2009,henig2009,sahai2013,robinson2011,palaniyappan2015,gonzalez-izquierdo2016,brantov2016,grassi2016,bychenkov2017}}.

Here we are interested in the role that RSIT may play in 
laser radiation pressure acceleration of ions that has
recently attracted a lot of attention~\cite{henig2009,naumova_PRL_2009,schlegel_POP_2009,grech2011,bin2015}.}
Indeed, when the plasma is opaque (for large enough plasma densities), 
and for thick enough targets, the so-called 
laser-driven hole-boring (\hb) regime occurs~\cite{macchi2005,klimo_PRSTAB_2008,robinson_NJP_2008,naumova_PRL_2009,schlegel_POP_2009,yan_PRL_2008,weng2012}.
Ions are accelerated in the 
electrostatic field induced by charge separation and a double layer structure known as a {\em laser piston} 
is formed, \reffig{f:intro}(c). 
{For non-relativistic ions, the resulting} ion energy scales as $\Ehb{}\propto{}a_0^2/\overline{n}_0$, 
where $\overline{n}_0$ is the normalized electron plasma density,
and thus there has been considerable interest in operating \hb\ as close to
the threshold density for \rsit\ as possible\rf{macchi2010,robinson2011,weng2012,mironov2012}.

In this work we show that the transition from the
\rsit\ to the \hb\ regime is associated with a much richer dynamical behavior
than previously {reported}, owing to the complex interplay of fast electron generation 
and ion motion. {In order to characterize the regime of interaction we perform a parametric {scan} in the $a_0$-$n_0$
plane and study signatures of \rsit\ in \refsect{s:detection}.}
{In contrast to previous studies~\cite{weng2012}, which characterize the regime of interaction in the asymptotic, long time limit,
we do consider the full time evolution, including {transient dynamics}.
This is particularly important in the mobile ion case and it allows us to uncover} a new \emph{dynamic transition} regime in which
the transition from \rsit\ to \hb\ occurs dynamically, \ie\ during the course of interaction. 
In order to understand the exact mechanism we develop a dynamical systems description
based on {the effect of ion motion on} electron phase-space separatrices in \refsect{s:sep}. {It is shown that the time scale over which ion becomes important is much shorter than usually anticipated leading to a dependence of the transition threshold on the ion charge-to-mass ratio.}
Moreover, the {dynamic transition} regime is shown to strongly depend on kinetic effects developing in the early stage of interaction and
can be controlled by varying the temporal profile of the laser pulse. 
{The importance of studying transient effects is emphasized by
comparing ion spectra in the conventional near-critical \hb\ regime and the 
dynamic transition regime in \refsect{s:ionSpec}. 
In the latter case much smaller energy dispersion is observed.
{Finally, in Sec.~\ref{s:discuss}, we  discuss the differences of the \transient\ regime 
with some previously explored near-critical regimes of ion acceleration~\cite{robinson2011,weng2012,sahai2013},
and present our conclusions.}
}

\section{\label{s:detection}Detection of the transition threshold}

The transition from the \hb\ (opaque) regime to \rsit\ is investigated using 1D3P PIC
simulations performed with the code EPOCH~\cite{arber2015}. 
The $(a_0,\,n_0)$-parameter 
plane was scanned to locate the transition threshold $n_{\mathrm{th}}$ for
different values of the ion charge-to-mass ratio corresponding to hydrogen, helium and immobile ions,
\reffig{f:fluidPIC}. 
The simulation box extends from $x=-L$
up to $x=L$, where $L=200\lambda_L$. The plasma fills half of the box 
with a constant electron density $n_0$ {and a step-like plasma-vacuum interface}. 
The initial electron
and ion temperatures are $T_i=T_e=5\times10^{-4}{}m_e{}c^2$.
The plasma is irradiated by a CP laser pulse with normalized 
vector potential 
\begin{equation}\label{eq:incident}
  \g a_L(x,t)=\frac{a_0}{\sqrt{2}} f(t)\,[\hat{\g y}\cos\xi +\hat{\g z}\sin\xi], 
\end{equation}
where $\xi={\omega_L t-k_L x}$, $k_L=\omega_L/c$ and the envelope $f(t)$ is a flat-top profile with a $\sin^2$ ramp-up of duration $\tau_r$.  
The pulse reaches the plasma at $t=0$ and the total simulation time is
$t_{\rm  sim}=2L/c$.
The spatial resolution is set to $\Delta{}x=0.8\lambda_{D}$, where 
$\lambda_{D}=\sqrt{\epsilon_0{}T_e/e^2{}n_0}$ is the
Debye length of the unperturbed plasma, 
the time-step is $\Delta{}t=0.95\Delta{}x$
and 1000 macroparticles-per-cell have been used.  

{In order to determine the density threshold $n_{\mathrm{th}}$ between the two regimes of interaction, we
examine two different time-series which are associated to either the velocity of the pulse front or the 
overlap of the laser pulse with the plasma electrons.} 

First, the pulse front position $x_f(t)$ is identified as the largest solution
of $a(x_f, t)=a_0/2$~\cite{weng2012}, where $a(x,t)=e|\mathbf{A}(x,t)|/(m_ec)$
is the normalized amplitude of the vector potential, see~\reffig{f:intro}(b). The pulse front position
moves deeper into the plasma at a velocity $v_f$ that strongly depends on the interaction regime.
In the opaque regime,  which occurs for $n_0>n_{\mathrm{th}}$, propagation is dominated by transfer of momentum from the laser photons
to the ions and $v_f$ equals the so-called hole-boring (or piston) velocity~\cite{naumova_PRL_2009,robinson2009}
\begin{equation}
  v_{\rm HB}=c \beta_{0}/(1+\beta_{0})\,,
\end{equation}
where $\beta_{0}=a_0/\sqrt{2m_i{}n_{i0}/(m_e{}n_c)}$, 
$m_i$ is the ion mass and $n_{i0}$ is the ion plasma density.
{As outlined in the introduction, defining the \rsit\ regime is
not straightforward when boundaries are involved. Here,
we adapt the point of view of earlier works which associated the \rsit\
regime in the immobile ion case 
with the absence of a standing wave solution~\cite{cattani2000,goloviznin2000,eremin2010,siminos2012}.
In the \rsit\ regime with mobile ions 
no double layer (relativistic piston) is formed and transfer of momentum to 
ions is minimal. This operating definition of \rsit\
for plasmas with an interface with vacuum implies deviations from the relativistic 
dispersion relation applicable in plasmas of infinite extend~\cite{akhiezer1956,kaw1970}.
Eventhough the energy balance has been invoked in
a number of works in order to determine the front 
propagation velocity in the \rsit\ regime~\cite{guerin1996,robinson2011a,weng_POP_2012},
no generally valid, closed-form solution exists~\cite{weng_POP_2012,weng2012}.}
{Therefore, in order to determine if the laser-front velocity $v_f$ in mobile ion 
simulations corresponds to propagation in the \rsit\ regime we compare it 
with $v^{\infty}_{\mathrm{SIT}}$, the front velocity from immobile ion simulations with otherwise identical interaction parameters. 
For laser amplitudes in the range $5\leq a_0 \leq 25$ that we study here,
it is expected that $v^{\infty}_{\mathrm{SIT}}>v_{\rm HB}$~\cite{weng2012}.}
We thus anticipate that at the threshold density  
for the transition from \hb\ to \rsit\ a discontinuous change of $v_f$
occurs. 

The second quantity on which we rely to distinguish between the opaque and transparency regimes
provides a measure of the overlap of the laser pulse with plasma
electrons. It is the {\em cross-correlation function}
\begin{equation}
  \Ssit(t)=n_c^{-1}\lambda_L^{-1}\int_{-L/2}^{L/2}dx\,n_e(x,t)\,|a(x,t)|^2 
\end{equation}
introduced in \refref{mironov2012}. In the \hb\ regime the laser-pulse overlap with plasma electrons
is limited to the electron skin-depth~\cite{mironov2012,schlegel_POP_2009}, see \reffig{f:intro}(c), and therefore $S(t)$ is expected to remain approximately constant (and small)
during the interaction. On the other hand, in the \rsit\ regime we expect $\Ssit$ to increase linearly
with time as the laser-pulse propagates deeper into the plasma at the constant velocity $v_f$.

\begin{figure}[t!]
 \begin{center}
  \includegraphics[width=0.8\columnwidth]{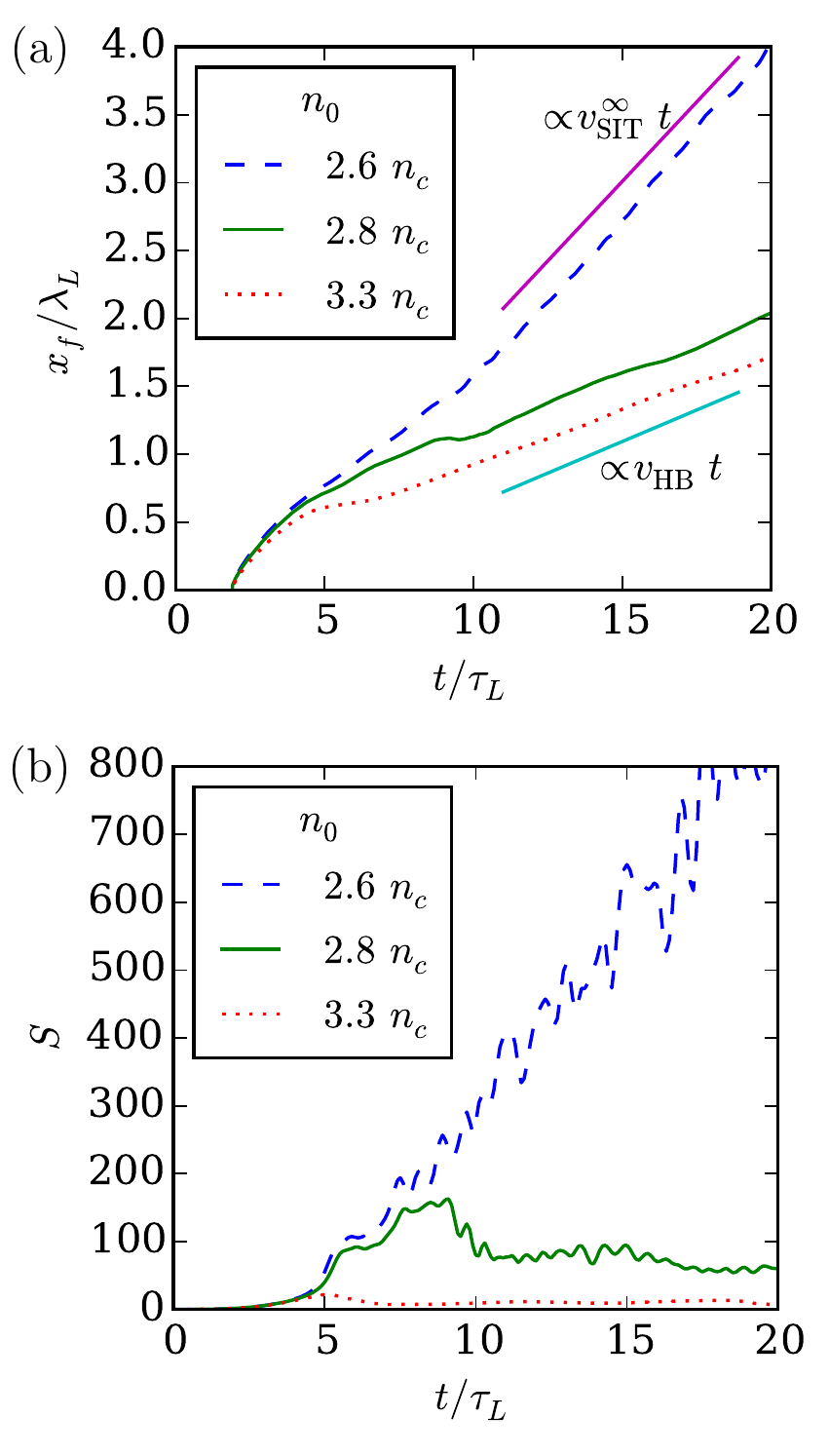}
  \caption{\label{f:compare_S_xf_t}
  (a) Pulse front position $x_f(t)$ for $a_0=10$, $\tr=4\tl$ and different densities, 
  $n_0=2.6,\,{}2.8$ and $3.3\,n_c$  (\rsit, \transient\ and \hb\ regime, respectively).
  The upper and lower straight solid lines correspond to front propagation 
  with $\vsit^{\infty}$ (with $n_0=2.6\,n_c$) 
  and \vhb\ (with $n_0=3.3n_c$), respectively. 
  (b) Cross-correlation function $\Ssit(t)$ for the simulations of panel (a).
  }
 \end{center}
\end{figure}

{With these two methods we 
can numerically determine the density threshold $n_{\mathrm{th}}(a_0)$ that {delineates} the HB ($n_0>n_{\mathrm{th}}$)
from the \rsit\ ($n_0<n_{\mathrm{th}}$) regime.}
We begin with 
the case of a hydrogen plasma and a pulse with $a_0=10$ and ramp-up time $\tau_r=4\tau_L$,
where $\tau_L=2\pi/\omega_L$.
In \reffig{f:compare_S_xf_t}(a) and \reffig{f:compare_S_xf_t}(b),
we plot as a function of time and for different $n_0$ the position of the
pulse front $x_f$ and the cross-correlation function $\Ssit$, respectively. 
For $n_0=3.3$, we observe that, after an
initial stage of duration $\simeq\tr$ during which a Doppler-shifted
standing wave~\cite{esirkepov2004} is formed, the front propagation velocity reaches a 
constant value $\vf=0.08\,c$. This matches very well the analytically predicted
hole-boring velocity 
$v_{\rm HB}=0.083\,c$. Moreover, 
$\Ssit$ remains approximately constant for $t>\tr$. 
This is characteristic for the \hb\ regime, {in which the overlap
of the laser pulse with plasma electrons is limited to the skin depth~\cite{mironov2012}}.

For $n_0=2.6n_c$ on the other hand,
the pulse front propagates with a velocity which 
at large times approaches the constant value $\vf=0.24\,c$,~\reffig{f:compare_S_xf_t}(a).
{This is much higher than $\vhb=0.09$ and very close to $v_f^{\infty}=0.23\,c$
obtained by performing a simulation with immobile ions and identical interaction
parameters.} This shows that this regime
of propagation is indeed dominated by electron motion effects.
In addition, $\Ssit$ increases approximately linearly after $t=\tr$.
{This implies that the laser overlap with plasma electrons increases with time as
expected in the \rsit\ regime~\cite{mironov2012}.}

For intermediate densities, 
between these two clearly defined regimes of propagation, we  observe a 
behavior that has not been identified before.  As an example, we show in \reffig{f:compare_S_xf_t}
the case $n_0=2.8n_c$ for which the pulse front propagates
initially with a velocity $\vf=0.11\,c$ larger than $\vhb=0.09\,c$ until up to approximately
$t\sim9\tl$. After this time the front velocity changes {abruptly} and matches
closely the \hb\ velocity.  The change in velocity between the initial
and final stages of propagation is subtle, and thus it is essential to also examine $\Ssit(t)$. 
In \reffig{f:compare_S_xf_t}(b) we
see that during the initial stage $\Ssit$ grows linearly, as is typical
of the \rsit\ regime. However, for $t>9\tau_L$ this growth saturates and
an almost constant value of \Ssit\ is reached, as is typical of the
\hb\ regime. This demonstrates the existence of a dynamic
transition from \rsit\ to \hb.

{In order to check the applicability of these results beyond the specific case studied so far, 
we performed a parametric scan for the transition threshold in 
the $(a_0$-$n_0)$ plane for immobile ions, helium and hydrogen. The results are summarized in \reffig{f:fluidPIC},
in which the width of the error-bar indicates the extent of the dynamic transition regime.
{We observe that RSIT occurs at much lower densities for mobile than for immobile ions.
Moreover, we see that the transition to RSIT occurs at lower density for ions with higher 
charge-to-mass ratio, as also 
observed in previous numerical simulations~\cite{weng2012}. 
We note that for mobile ions the transition occurs below the cold fluid theory threshold $n_{\mathrm{SW}}$ for existence
of a standing wave with immobile ions~\cite{cattani2000,goloviznin2000}, shown 
as a red solid curve in \reffig{f:intro}(b). These observations suggest that we need to} 
study the interplay of kinetic effects and ion motion in order to gain a qualitative understanding of the transition
mechanism, {a task that will be pursued in \refsect{s:sep}}.

\section{\label{s:sep}Importance of kinetic and finite ion mass effects}

\subsection{Phase-space separatrices}

\begin{figure*}[ht!]
  \includegraphics[width=\textwidth]{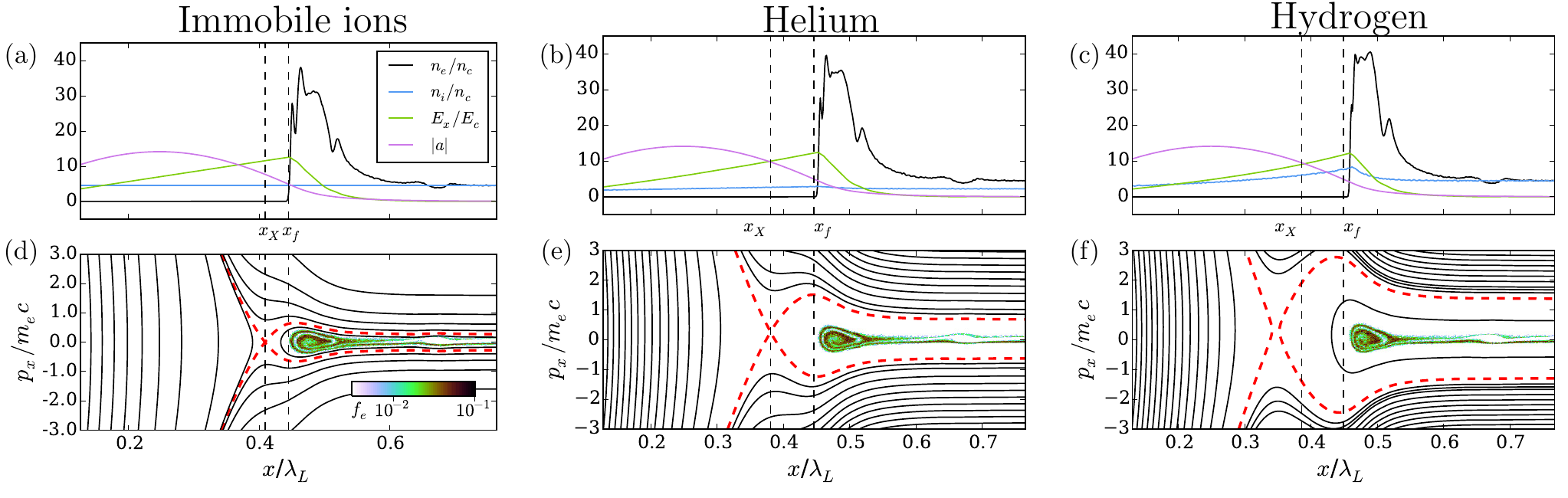}
  \caption{\label{f:hbsit_l_PS_ZM} 
  {Results of PIC simulations in the total reflection regime 
  ($n_0=4.5n_c$, $a_0=10$, $\tau_r=4\tl$) for different ion charge-to-mass
  ratio: (a,d) immobile ions, (b,e) helium, (c,f) hydrogen.
  Top panels show electron (black solid line) and ion density (blue solid line), electric field (green solid line) and vector potential amplitude (magenta solid line)
  and bottom panels show the electron distribution function $f_e(x,p_x,t)$ and contours
  of the Hamiltonian. The separatrices of bounded and unbounded electron motion
  are shown with red dashed line. All snapshots are shown for $t=5\tl$.}
  }
\end{figure*}

{As we will show, the transition to \rsit\
is in large part determined by laser energy absorption, which} 
in near-critical plasmas 
can be significant even with CP pulses\rf{guerin1996,ghizzo2007,robinson2011a,siminos2012}.
During the early stage of the interaction the ponderomotive 
force of the laser pulse 
accelerates electrons deeper into the plasma, {until it is shielded
by an electron density spike and wavebreaking occurs.}
Some of the accelerated electrons are trapped in the potential
well formed by the combination of the ponderomotive and 
electrostatic potentials.
{The exact mechanism of plasma heating is highly involved
and a detailed model is still lacking. Here we will show that we can gain insight
into kinetic effects despite the lack of a model of electron 
heating by using topological information encoded in distinguished trajectories in single-electron phase-space.}
In the case of immobile ions the escape of electrons
from single-particle separatrices at the plasma-vacuum interface was shown to be
responsible for transition to \rsit~\cite{siminos2012}. 
In particular, it was demonstrated that the width of
these separatrices decreases with decreasing density {$n_0$. Below a certain density $n_0$, finite amplitude perturbations in longitudinal
momentum $p_x$ can then lead to electron escape to the vacuum, lowering the electrostatic field. Then the ponderomotive
force prevails and pushes the electron front deeper in the target. This cycle repeats allowing laser pulse penetration in the target.} 

For the case of mobile ions the situation is more involved since
{the transient nature of ion motion during the early stages of the interaction
implies that well-defined separatrices may not exist.
In order to make progress we assume that such separatrices 
between escaping and confined trajectories
do exist over the electron time-scale and verify this assumption {\emph{a posteriori}}.
In particular,} we transform the
single-electron Hamiltonian
\begin{equation}\label{eq:H}
  H(x,p_x,t)=m_e c^2\,\sqrt{1+a(x,t)^2+p_x^2/m_e^2c^2}-e\phi(x,t)
\end{equation}  
to a frame moving with the instantaneous front velocity $\vf$. 
Here, $\phi(x,t)$ is the
{instantaneous scalar (electrostatic)} potential and 
$p_x$ is the electron momentum.
The Lorentz transformed Hamiltonian reads
\begin{equation}
H'=\gamma_f\left[H-v_f{}p_x\right],
\end{equation}
with $\gamma_f=(1-v_f^2/c^2)^{-1/2}$ (where a prime denotes 
a Lorentz-transformed coordinate). The
potentials and front velocity $\vf$ are determined from our PIC simulations.  
We assume that in the frame moving with velocity $\vf$ 
{a quasi-steady state of equilibrium between the ponderomotive 
and electrostatic force has been reached. In particular, we assume that}
the variation of the potentials
due to ion motion is slow compared to the typical time-scale for
electron motion and thus, $H'$ can be treated as time-independent.
{Although we plot contours of $H'$ both in and out of the plasma, we are interested in
their form in the charge separation layer, where fast electron dynamics have small effect on the fields~\cite{siminos2012}.}

{Separatrices are associated with saddle type (unstable) equilibria of the
equations of motion {(referred to as X-points)}.  
Taking into account Hamilton's equations, the equilibrium condition is written
$\dot{x}'=\partial H'/\partial
p'_x=0,\, \dot{p}_x'=-\partial H'/\partial x'=0$. 
The separatrices for electron motion are determined as iso-contours of $H'$
associated with its local minima.
Distinguishing saddle (unstable) from center (neutrally-stable) equilibria would involve examining second
derivatives of $H'$. 
However, for our purposes, the distinction will be
clear by inspection of phase-space plots.}
{Examples of separatrices are plotted in \reffig{f:hbsit_l_PS_ZM}, which will be discussed in detail in \refsect{s:ionZA}. 
{The critical momentum magnitude $|\pcr|$ is defined as the minimum momentum that an electron at the plasma boundary must 
have in order to escape to the vacuum. In the immobile ion case it is equal to the momentum gained by an electron placed at 
(the vicinity of) the X-point when it crosses the plasma boundary $x_b$~\cite{siminos2012}.}
}

\subsection{\label{s:ionZA}Effect of ion charge-to-mass ratio: a case study}

{We begin by examining the effect of ion charge-to-mass ratio on
the dynamics in the total-reflection regime.
In \reffig{f:hbsit_l_PS_ZM} we show the electron phase space
from simulations for $n_0=4.5\,n_c$, $a_0=10$, $\tau_r=4\,\tl$
for the cases of hydrogen, helium and immobile ions. {These
parameters were chosen so that all three cases correspond to the opaque regime.}
We show snapshots at $t=5\tl>\tr$ so that {the flat-top part of the pulse has reached the target.}
We choose to compare the phase space at this early stage of interaction
because, as will become evident in the following, this is when the transition to \rsit\ is determined. We find that there are no significant differences in 
the width of the electron distribution close to the plasma-vacuum interface at this stage. This shows that any differences in electron heating due to 
laser pulse energy being expended in ion motion are minimal and cannot
explain the difference in transition threshold.}

{\refFig{f:hbsit_l_PS_ZM} allows us to confirm 
that the electrostatic field is perturbed (compared to the immobile ion case) 
due to ion motion already at this early stage. 
The reduction in the electrostatic field in the charge separation 
layer [more visible for hydrogen, \reffig{f:hbsit_l_PS_ZM}(c)] is larger at 
the position of the X-point $x_X$ rather than at the position of the electron front $x_f$.
This is due to the fact that the perturbation in ion density depends both on the 
magnitude of the electrostatic field and on the time over which it acts on ions.
Before ion motion becomes important, the field increases approximately linearly with $x$,
{
\begin{equation}\label{eq:ExEquil}
 E_x/E_c = \overline{n}_0\,k_L x\,\qquad 0<x<x_b\,,
\end{equation}
with $E_c=m_e c\, \omega_L/e$ the so-called Compton field.
}
On the other hand, since it takes a finite time 
for the charge separation layer to be setup, the time over which an ion is accelerated decreases with its initial position $x$. As a result, ions close 
to the plasma boundary $x_b\simeq x_f$ did not yet have enough time to respond and the difference in the position 
of the front $x_f$ between the mobile and immobile ion cases is negligible. On the other hand, the position of the X-point
is determined by the balance of the ponderomotive and electrostatic force.
Due to the reduction of the electrostatic field in the middle of the charge separation layer,
a new equilibrium is reached at a position where the magnitude of 
the ponderomotive force is smaller, \ie, the X-point $x_X$ is moved towards the left where 
the slope of $|a|$ is smaller, see \reffig{f:hbsit_l_PS_ZM}.
At the same time the magnitute of critical momentum for escape to vacuum 
(the separatrix width)
becomes larger as the distance of $x_X$ and $x_b$ increases.}
{
To understand this qualitatively, note that a test electron with small positive initial momentum placed at $x_X$ will 
gain a net momentum (approximately equal to the critical momentum magnitude $|\pcr|$) 
while moving up to $x_b$, since the ponderomotive force is larger than the electrostatic force for $x_X<x<x_b$.
In the mobile ion case the same electron would experience a larger 
average accelerating force (due to the reduction in electrostatic field) for a larger distance (due to the increase in $x_b-x_X$)
therefore gaining larger net momentum.
}

\subsection{Time-scale for ion motion}

\begin{figure*}[ht!]
  \includegraphics[width=\textwidth]{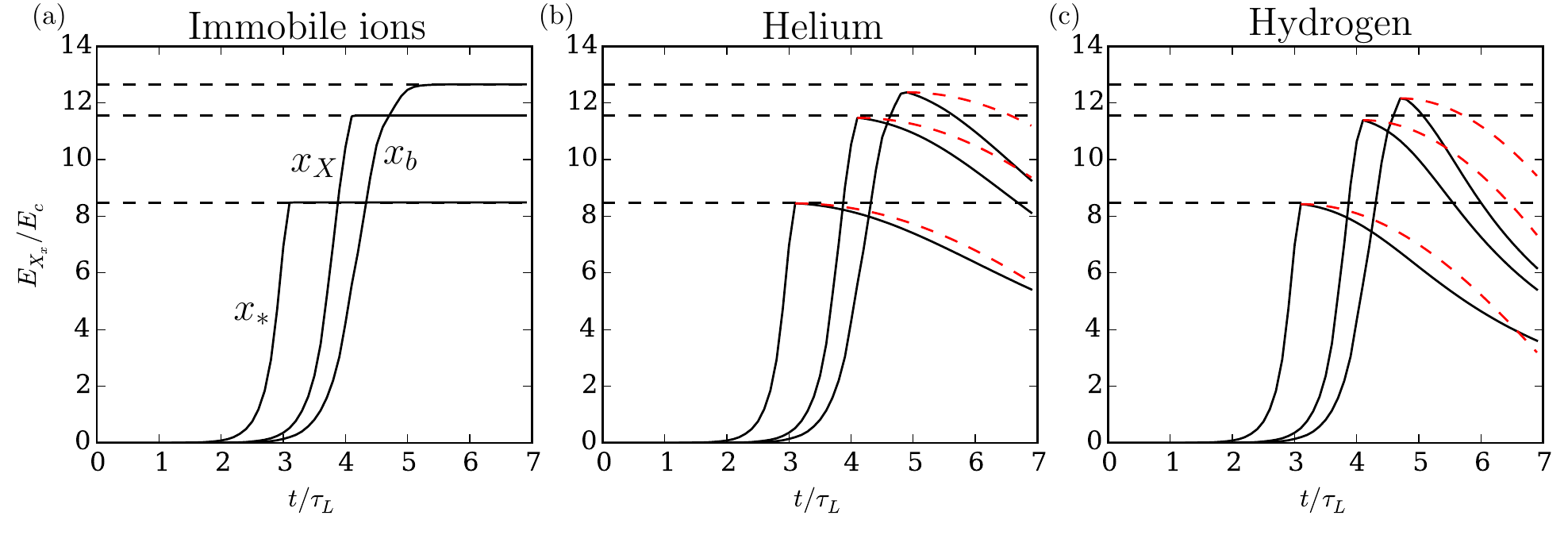}
  \caption{\label{f:hbsit_l_Ex_Change} 
  Evolution of the local electrostatic field $E_x$ for the PIC simulations of \reffig{f:hbsit_l_PS_ZM} ($n_0=4.5n_c$, $a_0=10$, $\tau_r=4\tl$)
  for three different positions in the charge separation layer $x_{*}=0.3$, the X-point $x_X$ and the electron density boundary $x_b$. Dashed horizontal lines indicate the theoretically predicted value for the maximum electrostatic field for each $x$, according to \refeq{eq:ExEquil}. The solution of \refeq{eq:solEx}
  for $E_x(x,t)$ for each case is shown as a red, dashed curve. The initial time $t_0$ corresponds to the time, for each $x$, for which the electrostatic field assumes its maximum value.
  }
\end{figure*}

{ {Let us now give an estimate for the time-scale over which ion motion becomes important in the sense that it can affect the electron dynamics close to the interface $x \sim x_f$.}
Naively, an estimate could be provided by $2\pi\omega_{pi}^{-1}$, 
where  $\omega_{pi}=\sqrt{Z^2 e^2 n_{0i}/\epsilon_0 m_i}$ is the ion plasma frequency, 
$n_{0i}=n_0/Z$ is the ion number density, $Z$ is the atomic number of the ion species, 
and $m_i$ is the ion mass. 
For a typical case of hydrogen with $n_{0i}=n_0=3$, 
we find $2\pi\omega_{pi}^{-1}\simeq25\tl$. 
This appears to be too large to affect the transition dynamics according to the results of
\reffig{f:compare_S_xf_t}.

The main problem with the above estimate is that it does not take into account 
that transient ion motion can occur in the strong electrostatic field of order $\sqrt{2}a_0 E_c$ set up by the laser pulse ponderomotive force
and, thus, it does not depend on the laser strength $a_0$.
This is particularly important here, since \reffig{f:hbsit_l_PS_ZM} shows that a relatively small change in the electrostatic field $E_x$ can lead to change in critical momentum for escape of the order of $m_e\,c$. 
{Indeed, as we are here investigating the effect of ion motion on the electron dynamics, w}e can anticipate that a change in electric field of the order of $E_c$ (the typical field for relativistic electron effects), could lead to qualitative changes in dynamics even if the maximum field is many orders of magnitude larger than this.
We will now develop a simple model for the transient ion response at the early stage of interaction in order to estimate the time required for a change in electric field of order $E_c$ to occur.
}

{In order to obtain an upper bound for the response time of the ions we
model the interaction as a two stage process~\cite{capdessus2013}. Initially the electrons are pushed by the ponderomotive force and a charge separation layer is formed. {The resulting electrostatic field is a linear function of the space coordinate $x$, as described by \refeq{eq:ExEquil}}. 
At a second stage, ions are accelerated in this electrostatic field.
Since it takes a finite time to setup the charge separation layer, ions with smaller $x$
are accelerated for a longer time (but experience a smaller electric field).
Treating the ions as a cold fluid, we write the ion momentum equation as
\begin{equation}\label{eq:ionMomentum}
	m_i\,n_i\frac{\partial V_i}{\partial t} + m_i\,n_i\, V_i\frac{\partial V_i}{\partial x} = q_i n_i E_x(x,t)\,,
\end{equation}
where $V_i(x,t)$ is the ion fluid velocity. We let $t_0=t_0(x)$ denote the time at which the charge separation ``front" sweeps point $x$, 
and the field takes the value predicted by \refeq{eq:ExEquil}, \ie, the plateau in \reffig{f:hbsit_l_Ex_Change}(a) is reached. 
The ions are assumed initially at rest, $V_i(x,t_0)=0$, and we consider short enough evolution times 
that we may linearize \refeq{eq:ionMomentum} and drop the term  $V_i\partial_x V_i$. 
For the same reason we also ignore relativistic ion effects. 
Even though the ions obtain finite momentum at early times, their density response is expected to be minimal and, 
since we are only interested in obtaining an upper bound {on the characteristic time for ion motion to affect the electron dynamics}, 
the effect of ion density variations in the electric field (through Poisson's equation) will not be considered. 
Under these assumptions, we only need the longitudinal component of Maxwell-Ampere's equation to close the model,
\begin{equation}\label{eq:Ampere}
	j_x = -\epsilon_0 \frac{\partial E_x}{\partial t}\,.
\end{equation}
In the charge separation layer there are no electrons, so that $j_x=q_i n_i V_i$. Substituting \refeq{eq:Ampere} in the linearized version of \refeq{eq:ionMomentum} we obtain
\begin{equation}\label{eq:Vi}
	\frac{\partial^2 V_i}{\partial t^2}=-\omega_{pi}^2 V_i\,.
\end{equation}
This has the solution
\begin{equation}\label{eq:solVi}
	V_i(x,t)=\frac{q_i}{m_i\omega_{pi}}E_x(x,t_0)\sin\left[\omega_{pi}\left(t-t_0\right)\right]
\end{equation}
where for each $x$, $E(x,t_0)=\overline{n}_0 k_L E_c x$ from \refeq{eq:ExEquil} is taken as initial condition and  
we have enforced consistency of $\left.\partial_t V_i\right|_{t=t_0}$ with \refeq{eq:ionMomentum}. 
Taking into account \refeq{eq:solVi},  the solution of \refEq{eq:Ampere} can be written
\begin{equation}\label{eq:solEx}
	E_x(x,t)=E(x,t_0)\cos\left[\omega_{pi}\left(t-t_0\right)\right]\,.
\end{equation}
For $a_0\gg a_b\gg 1$ 
we have that~\cite{siminos2012} $k_L x_b\simeq \sqrt{2}a_0/(\overline{n}_0)$ and thus $E_x(x_b,t_0)\simeq\sqrt{2}a_0 E_c$.
From \refeq{eq:solEx} we find that an $\mathcal{O}(E_c)$ change in the electric field at $x_b$, $\Delta E_x\simeq-E_c$, occurs at a timescale
\begin{equation}\label{eq:TiMod}
	\tau_i=\frac{g(a_0)}{\omega_{pi}}=g(a_0)\sqrt{\frac{A\, n_c\,m_p}{Z\,n_0\,m_e}}\tau_L\,,
\end{equation}
where 
\[
	g(a_0)=\arccos\left(1-\frac{1}{\sqrt{2}a_0}\right)\,.
\]

We note that \refeq{eq:TiMod} is derived under the assumption of a large initial electric field. It is valid only in the limit $a_0\gg a_b\gg 1$, and becomes singular for $a_0<1/(2\sqrt{2})$.
For completeness, we mention that in the large density regime, $n_0\gg a_0^2$, the maximum electrostatic field at $x_b$ scales as~\cite{siminos2012} $E_{x,\mathrm{max}}/E_c\simeq 2a_0^2/\overline{n}_0^{1/2}$ and a different limiting behavior can be derived, $g(a_0)=\arccos\left(1-\overline{n}_0^{1/2}a_0^{-2}/2\right)$.

\refFig{f:hbsit_l_Ex_Change} shows the evolution of the electrostatic field in the initial phase of the interaction for the three cases of  \reffig{f:hbsit_l_PS_ZM} and for three different positions in the charge separation layer $x_{*}=0.3$, the X-point $x_X$ and the electron density boundary $x_b$. We see that the solution of \refeq{eq:solEx}
for $E_x(x,t)$ (red, dashed curve) agrees well with the simulations for smaller $x$, while for larger $x$ the observed change in $E_x$ is faster than predicted by \refeq{eq:solEx}. This is because we did not take into account the fact that ions will start to move even before the field reaches the value predicted by \refeq{eq:ExEquil}. Indeed, as seen in \reffig{f:hbsit_l_Ex_Change}(b) and (c), the time interval during which the electric field rises is finite and increases with the position $x$. Although we could, in principle, account for this by solving \refeq{eq:Vi}  with initial condition $V_i(x,t_0)\neq0$, we will not pursue this here since we are only interested in obtaining an estimate.
Moreover, in the above derivation, we did not take into account the effect of ion density variation. At later times, this leads to deviation from the sinusoidal behavior predicted by \refeq{eq:solEx}. However, even in the worse case scenario of \reffig{f:hbsit_l_Ex_Change}(c) this only occurs after a change of order $E_c$ in $E_x$ has taken place. Therefore, \refeq{eq:TiMod} constitutes a useful upper bound for the time-scale at which ion motion becomes important in our problem. For the case of helium (hydrogen) with $n_0=4.5n_c$ and $a_0=10$, \refeq{eq:TiMod} predicts
a change in electric field of the order of $E_c$ at time 
$\tau_i=1.7\tau_L$
($\tau_i=1.2\tau_L$) after the time $t_0=4.1\tau_L$ at which the charge separation layer has been set up at the X-point (found from the PIC simulations, see \reffig{f:hbsit_l_Ex_Change}). Although this is still a conservative upper bound, it matches much better the results of \reffig{f:hbsit_l_PS_ZM}(b) and (c) than the naive scaling $2\pi\omega_{pi}^{-1}=28\,\tau_L$ and $20\,\tau_L$ obtained for helium and hydrogen, respectively.
}

\subsection{Transition to RSIT}

{In order to establish the connection of the separatrix width to
the transition to \rsit, we now concentrate in the case of a hydrogen plasma 
and reduce the density, compared to \reffig{f:hbsit_l_PS_ZM}(c),
to the lowest possible density $n_0=3.3\,n_c$ in the \hb\ regime.}
In \reffig{f:PS_sep_hb} we show, for two different times $t=5\tau_L$ and $t=15\tau_L$, 
the results of a simulation with $a_0=10$, $n_0=3.3\,n_c$.
{For these parameters cold fluid theory with immobile ions predicts
that no standing wave solution exists [\reffig{f:fluidPIC}] and electrons from the
dense electron layer would be able to escape to the vaccuum leading to \rsit\ according to 
the scenario in~\refref{eremin2010,siminos2012}.
However, we see in \reffig{f:PS_sep_hb}(c) that due to ion motion
}
a separatrix merely wide enough that no electrons escape
to the vacuum during the initial stage of the interaction exists.
{The separatrix width is smaller than in the case $n_0=4.5n_c$
of \reffig{f:hbsit_l_PS_ZM} because it takes longer for ion effects to become important
in this case of lower density (according to \refeq{eq:TiMod}, $\tau_i=1.4\tau_L$)}.
With time, a double layer is
formed and propagates deeper into the plasma as a laser
piston {[\reffig{f:PS_sep_hb}(b,d)].  At this stage} 
the separatrix becomes wider {in $p_x$} as ions catch
up with the electrons, 
reducing the charge separation induced electrostatic field. This contributes 
to the stability of the \hb\ process {as electrons cannot escape at this point}.

\begin{figure*}[ht!]
  \includegraphics[width=0.8\textwidth]{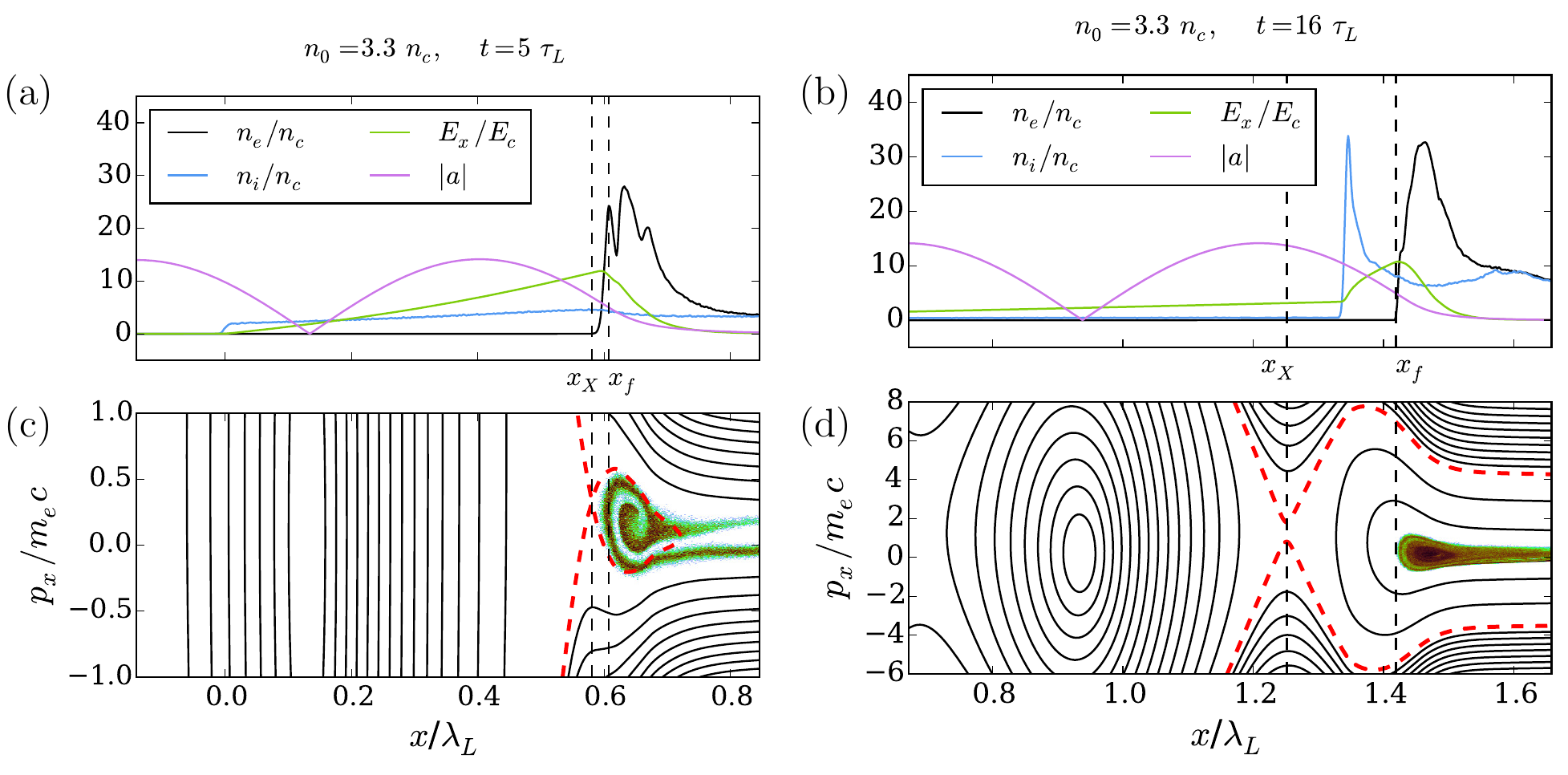}
  \caption{\label{f:PS_sep_hb} 
  \hb\ regime simulation ($n_0=3.3n_c$, $a_0=10$, $\tr=4$, hydrogen): Electron and ion density, electric field and vector potential amplitude 
  for (a) $t=5\tl$ and (b) $t=15\tl$. Electron distribution fuction 
  $f(x,p_x,t)$ 
  and iso-contours of $H'$ (black, solid lines), including the separatrices (red, dashed lines)
  in the lab frame for (c) $t=5\tl$ and (d) $t=16\tl$.
  }
\end{figure*}

We next examine typical dynamics in the \rsit\ regime, \ie, 
at lower density [$a_0=10$, $n_0=2.6\,n_c$, \reffig{f:PS_sep_rsit}]. 
{Lowering the density further decreases the effect of ion motion, $\tau_i\simeq1.6$, 
preventing the plasma to reach a quasi-static state which could trap electrons.
As electrons escape the space-charge is largely reduced
and the ions remain essentially immobile during the course of the simulation.
Therefore, the immobile ion results apply: the interaction is in the \rsit\ regime since 
$n_0<n_{\mathrm{SW}}$~\cite{cattani2000,goloviznin2000,eremin2010,siminos2012}.
Since the quasi-static approximation does not hold in this case, we do not plot separatrices in \reffig{f:PS_sep_rsit}. {However, we note that the fact that electron escape in the PIC simulations occurs for all values below $n_0=3.3$ for which the separatrix [\reffig{f:PS_sep_hb}(c)] was marginally wide enough 
to prevent electron escape justifies using the Lorentz-transformed Hamiltonian in order to define separatrices of confined and escaping electrons.}}
{We note that laser propagation in this \rsit\ regime is not
associated to the destruction of the electron density peak; the latter remains higher than 
the threshold $\nceff$ predicted by \refeq{eq:nceff}, see \reffig{f:PS_sep_rsit}(b).
Rather, while some electrons are pushed into the plasma, other electrons continuously escape in the region 
where they interact with the laser-pulse through a mechanism akin to beatwave heating~\cite{ghizzo2007}.
{We thus conclude that, as in the case of immobile ions~\cite{siminos2012}, electron escape drives transition to \rsit.}

\begin{figure*}[ht!]
  \includegraphics[width=0.8\textwidth]{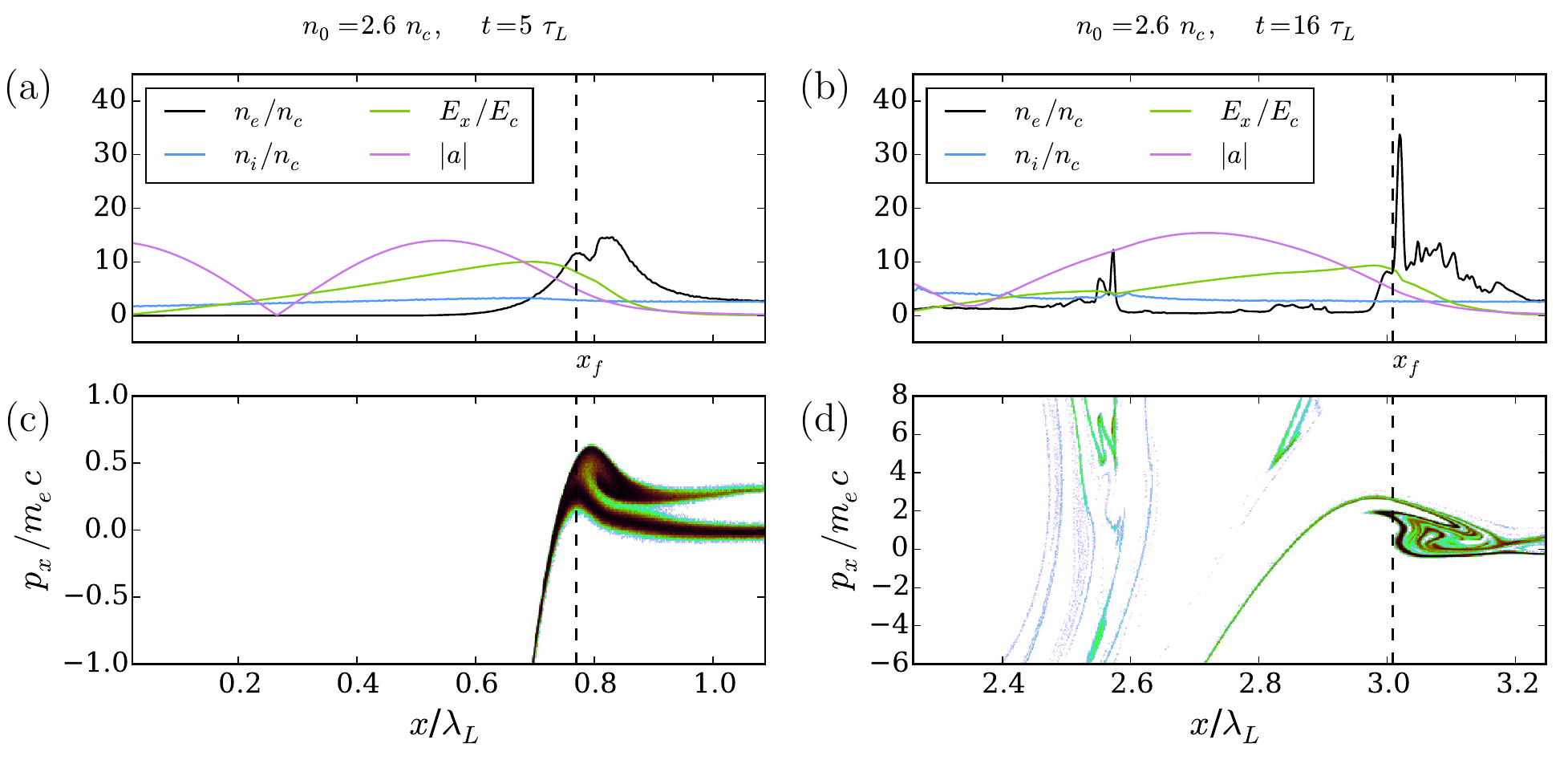}
  \caption{\label{f:PS_sep_rsit} 
  \rsit\ regime simulation ($n_0=2.6n_c$, $a_0=10$, $\tr=4$, hydrogen): Electron and ion density, electric field and vector potential amplitude 
  for (a) $t=5\tl$ and (b) $t=15\tl$. Electron distribution function 
  $f(x,p_x,t)$ 
  and, when applicable, iso-contours of $H'$ (black, solid lines), including the separatrices (red, dashed lines)
  in the lab frame for (c) $t=5\tl$ and (d) $t=16\tl$.
  }
\end{figure*}

{
For intermediate densities $2.7<n_0/n_c<3.3$ between the hole boring and \rsit\ regimes we find the dynamic transition regime. 
As an example we see in \reffig{f:PS_sep_transient} that for 
$a_0=10$, $n_0=2.8n_c$ electrons are initially escaping [panels (a,c)].
The estimate for the ion response time, $\tau_i\simeq1.5$, is slightly smaller than in the \rsit\ case, 
while at the same time the \rsit\ velocity $\vsit^{\infty}$ decreases with the density~\cite{siminos2012,weng2012}.
Therefore ions in the charge separation layer gain enough momentum to catch up with the electron front.
This leads to the eventual formation of a piston and of a potential well in which electrons are trapped [panels (b,d)]. 
Electron escape then saturates and the subsequent dynamics are of the \hb\ type.

For completeness, we note that for even larger laser
field amplitudes ($a_0\geq20$), interaction in the
\transient\ regime can be even more complex and a transition may
also occur in the reverse direction, from \hb\ to \rsit, 
since electrons accelerated by the beatwave heating mechanism~\cite{ghizzo2007}
can re-enter the plasma and destabilize the relativistic piston.
}

\begin{figure*}[ht!]
  \includegraphics[width=0.8\textwidth]{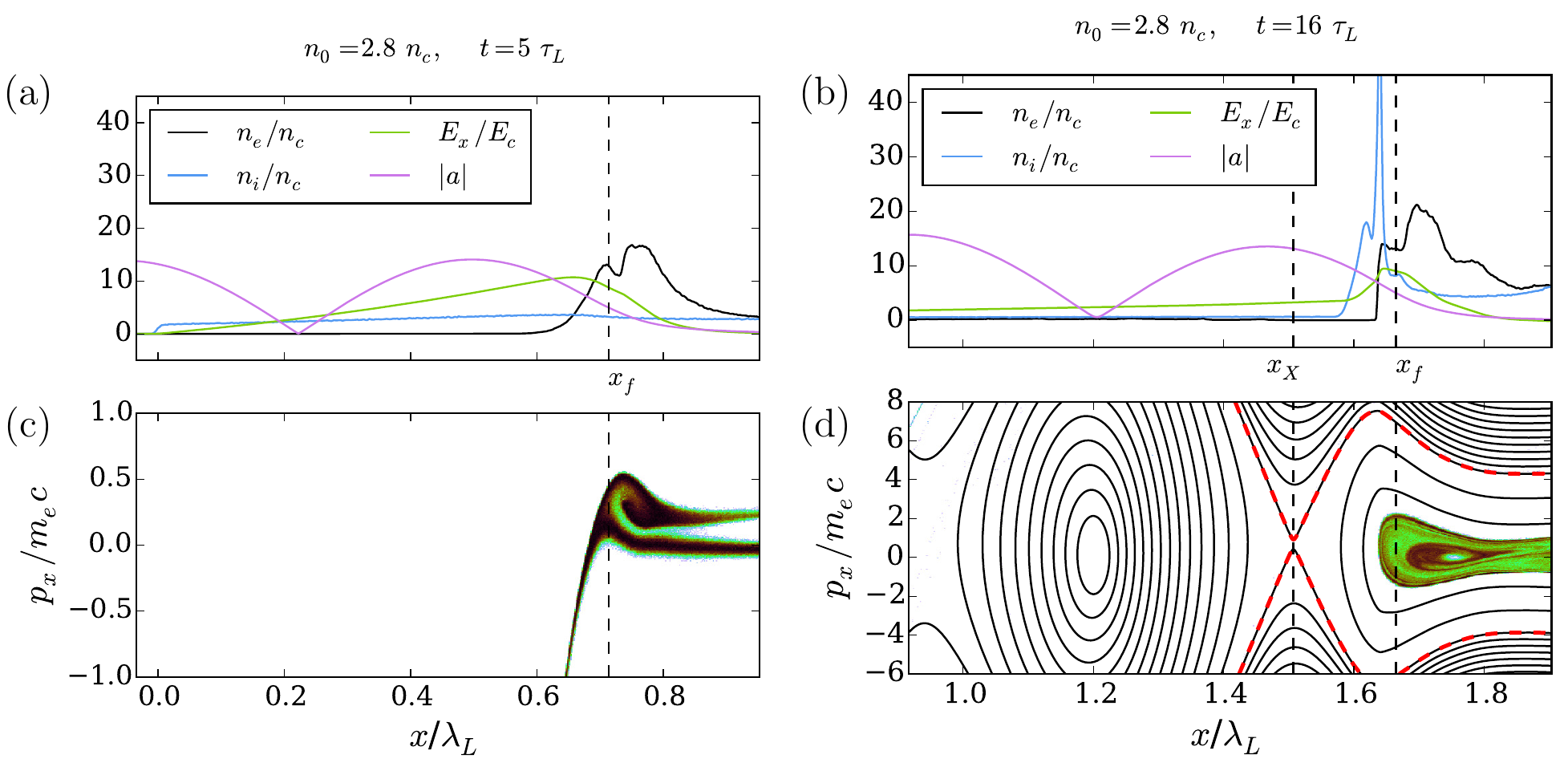}
  \caption{\label{f:PS_sep_transient} 
  Dynamic transition regime simulation ($n_0=2.8n_c$, $a_0=10$, $\tr=4$, hydrogen): Electron and ion density, electric field and vector potential amplitude 
  for (a) $t=5\tl$ and (b) $t=15\tl$. Electron distribution fuction 
  $f(x,p_x,t)$ 
  and iso-contours of $H'$ (black, solid lines), including the separatrices (red, dashed lines)
  in the lab frame for (c) $t=5\tl$ and (d) $t=16\tl$.  
  }
\end{figure*}

\section{\label{s:ionSpec}Effect of laser envelope on the transition threshold and ion energy distribution}

\subsection{\label{s:ionSpecA}Effect on the transition threshold}
Since kinetic effects in the early phase of interaction 
play an important role in the transition between
the different regimes, we can, to some extent, control the transition
by varying the shape of the laser pulse. The ponderomotive force
associated to a pulse with a shorter rise-time is larger than 
for one with a longer rise-time and this is expected to lead to stronger
electron heating {in the former case}~\cite{robinson2011a}. 
In order to illustrate this, we choose fixed values
of $a_0=10$ and $n_0=2.7n_c$ and perform simulations with different
pulse rise-times, $\tr=4,\, 7$ and $12\tl$. 

In \reffig{f:compare_S_xf_t_env} we show that for the shortest value {$\tr=4\tl$}
the pulse propagates in the \rsit\ regime, while as $\tr$ increases to $\tr=7\tl$
and $\tr=12\tl$, the \transient\ and \hb\ regimes are reached, respectively.  The relation
of this effect to electron heating is illustrated in
\reffig{f:spectra_tr}(a), where {the electron spectra are compared
at an early interaction time, $t=4.4\,\tl$, before electrons escape in any of these cases}. 
We find that electron
spectra in the case of shorter rise-time are broader than for longer
rise-times, {showing that electron heating indeed occurs at a higher rate for the pulse 
with shorter rise time $\tr$}. Moreover, it was verified by plotting the
electron separatrices (not shown) that the transition mechanism
is identical to the one described above. {In the case of $\tr=4$ 
the stronger electron heating leads to electron escape and triggers \rsit. For $\tr=12$ no 
electrons gain enough momentum to escape to the vacuum and we have \hb. Finally,
for $\tr=7$ some electrons escape but eventually ion response leads to a dynamic transition 
to \hb.}

\begin{figure}
 \begin{center}
  \includegraphics[width=0.8\columnwidth]{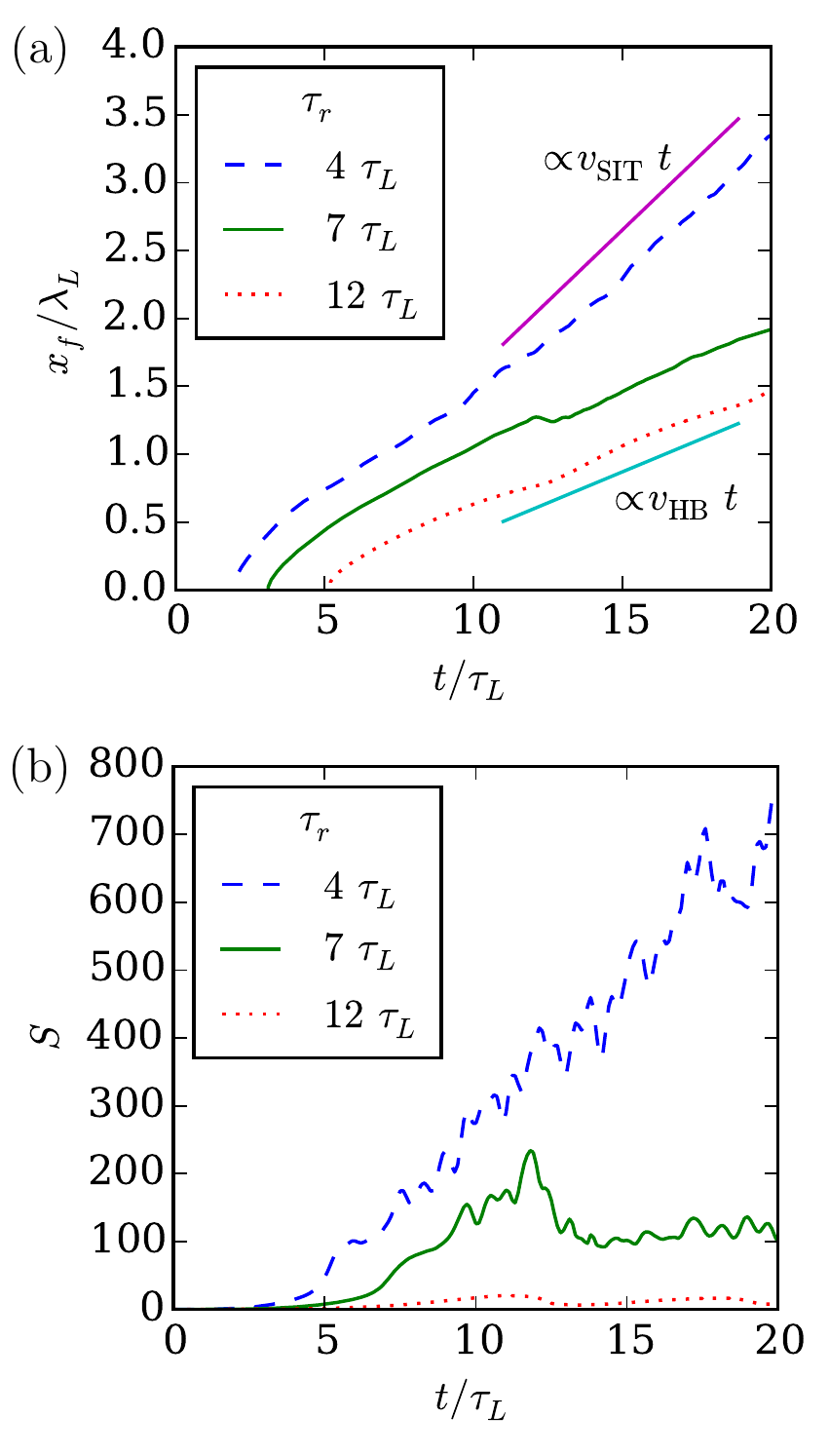}
  \caption{\label{f:compare_S_xf_t_env} (a) Pulse front position
    $x_f(t)$ for hydrogen plasma, $a_0=10$, $n_0=2.7n_c$ and
    different rise-times $\tr=4,\, 7$ and $12\,\tl$ 
    (\rsit, \transient\ and \hb\ regime, respectively). 
    (b) $S(t)$ for the same simulations.   }
 \end{center}
\end{figure}

\begin{figure}
  \includegraphics[width=0.8\columnwidth]{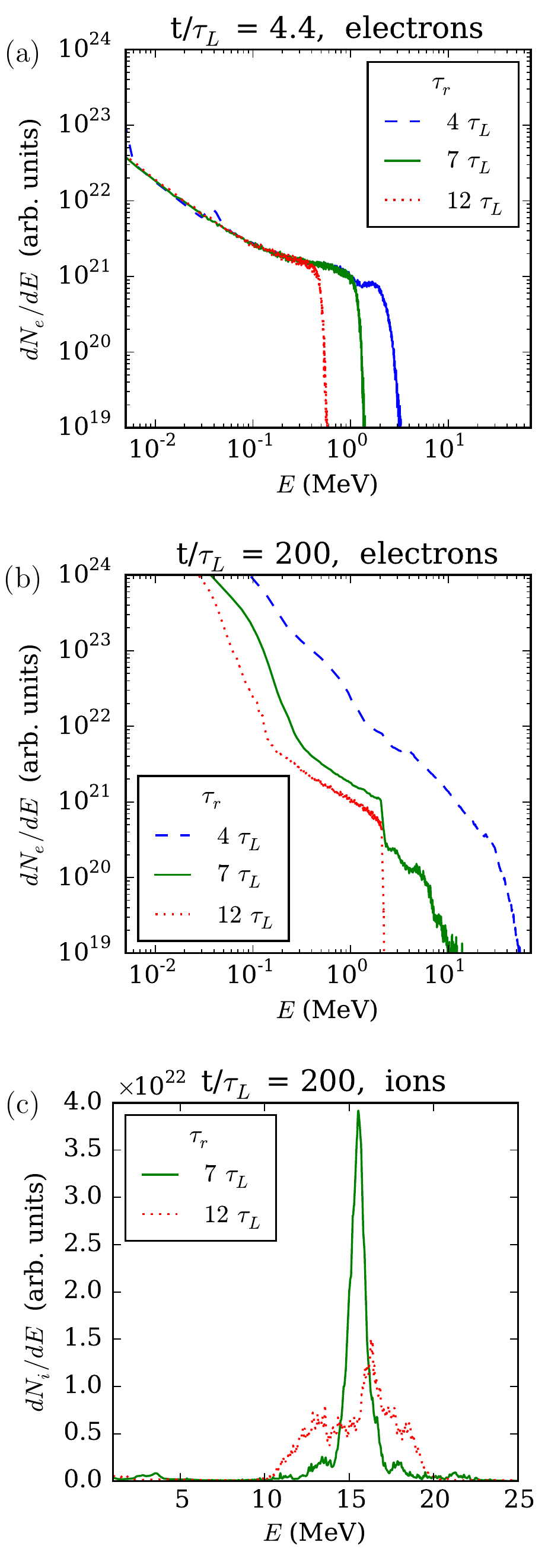}
  \caption{\label{f:spectra_tr} Energy spectra for 
  $a_0=10$, $n_0=2.7\,n_c$ and 
  different rise-times $\tr=4\,\tl$ (\rsit), $\tr=7\,\tl$ (\transient\ regime) 
  and $\tr=12\tl$ (\hb). (a) Electron spectra for 
  electrons with $x>\xf$ obtained at $t=4.4\,\tl$, (b) same as above but for at $t=200\,\tl$ 
  and (c) spectra for ions with $p_x>0$ obtained at $t=200\,\tl$.
  }
\end{figure}

\subsection{\label{s:ionSpecB}Effect on the ion energy distribution}

{In either the case of $\tr=7\,\tl$ (\transient) or $\tr=12\,\tl$ (\hb)
the long time dynamics corresponds to hole-boring. It is therefore worth 
asking whether there are any differences in ion spectra in these cases.
The ion spectra at $t=200\,\tl$ are shown in \reffig{f:spectra_tr}(c).
We observe that in the \hb\ regime ($\tr=12\tl$) the spectrum has a multi-peak
structure around the hole-boring energy $\Ehb=15.8$~MeV. }
{By contrast, for a typical simulation in the \transient\ regime ($\tr=7\, \tl$ and all other parameters kept unchanged),
we see in \reffig{f:spectra_tr}(c) that the spectrum has a much lower energy spread.
The {peak} energy $\mathcal{E}\simeq15.6$~MeV
is very close to the analytical prediction for \hb,
$\Ehb=15.8$~MeV, and the energy spread  (1 MeV or $6\%$ FWHM) 
is much smaller than in the pure \hb\ regime for $\tr=12\tl$ (correspondingly, $\mathcal{E}\simeq16.3$~MeV, 
and energy spread of $\simeq 5$~MeV or $30\%$ FWHM). }

{To explain the differences in the ion spectra, one has to examine into more detail the
dynamics of the double layer structure.  
Indeed, the broadening of the spectrum in the conventional \hb\ regime is
usually attributed to large amplitude periodic oscillations of the double layer,
known as piston oscillations~\rf{robinson2009,schlegel_POP_2009,grech2011}. 
These oscillations are illustrated in \reffig{f:piston_osc}(a) [for the conventional \hb\ case ($\tr=12\tl$)], 
where large scale ($\Delta E_{x,{\rm max}}/E_{x,{\rm max}} \simeq 0.3$) 
periodic fluctuations are observed in the temporal evolution of the maximum value electrostatic field.
These oscillations result in ions being reflected at different phases of the oscillating piston 
and therefore accelerated to different energies as described in \refref{robinson2009},
thus leading to ion bunching and modulation of the ion beam in $x-p_x$ phase space, 
sometimes referred to as 'rib-cage' structure, and illustrated in \reffig{f:piston_osc_PS}(b).   

Although the exact mechanism behind these oscillations is still largely not understood, 
e.g. no model yet describes the time at which they set in nor why they appear,
one can still get a deeper understanding of how they proceed by examining more closely
the time evolution of the maximum values of the electrostatic field and electron/ion densities, 
as shown in \reffig{f:piston_osc}(a-c).

The following discussion focuses on non-relativistic piston velocities, and 
builds on the previous analysis of piston oscillations as a three-step process
presented in Ref.~\cite{schlegel_POP_2009},
where the piston structure was also described within the framework of stationary
cold fluid theory.

In a first stage [region I in \reffig{f:piston_osc}(a-c)], an ion bunch is formed
in the charge separation close to the laser front, associated with an increase 
of the maximum ion density as shown in \reffig{f:piston_osc}(b). {This can 
be seen more clearly in the Supplemental Movie~1~\rf{hbsitSuppl}}.

In a second stage [region II in \reffig{f:piston_osc}(a-c)], this ion bunch crosses the charge separation layer and
is launched into the plasma. This results in the abrupt decrease of the electrostatic field  evidenced in \reffig{f:piston_osc}(a). 
Note that stage I and II are characterised by the maxima of ion and electron density as well as electric field being in a very close 
vicinity (Supplemental Movie~1~\rf{hbsitSuppl}). Furthermore, ion bunches launched into the target during the second stage 
have a velocity $\sim 2 v_{\mathrm{HB}}$. This can be seen in \reffig{f:piston_osc}(c) where the velocity computed from 
the position of the maximum ion density is about twice that computed from the maximum electron density moving at $v_{\mathrm{HB}}$ 
(note that the discontinuity in the position of the maximum ion density occurs when the ion bunch launched into the target 
becomes more dense than the the ion density peak in the charge separation layer and vice versa).

The characteristic time for these two first stages is related to the thickness $\Delta_e$ of the compressed electron layer.
As shown in Ref.~\cite{schlegel_POP_2009},  $\Delta_e \simeq c/\omega_{pe}$, and the duration of these first two stages is negligible with
respect to the characteristic time of the piston oscillations.

Of particular importance is the third stage [region III in \reffig{f:piston_osc}(a-c)], during which not yet reflected 
ions move deeper into the charge separation layer, 
thus increasing the charge imbalance and enhancing the electrostric field as observed in \reffig{f:piston_osc}(a). 
The rate of increase of the electrostatic field can be estimated from  Amp\`ere's equation
as $dE_x/dt \sim Z e n_{i0} v_{\mathrm{HB}}/\epsilon_0$, and the characteristic duration of this stage is $\tau_3 \simeq 2\Delta_i/v_{\mathrm{HB}}$, 
where $\Delta_i$ is the width of the charge separation layer. 
The latter can be estimated from the piston model proposed in Ref.~\cite{schlegel_POP_2009} as
$\Delta_i \simeq v_{\mathrm{HB}}/(3\,\omega_{pi})$, for $v_{\mathrm{HB}} \ll c$. 
This leads $\tau_3 \sim 2/(3\omega_{pi})$, much larger than the characteristic duration of the first two stages ($\propto \omega_{pe}^{-1}$) so that the characteristic duration of an oscillation is $\tau_{\rm osc} \sim \tau_3 \simeq \omega_{pi}^{-1}$.
The total increase of the electrostatic field during this stage can then be computed as 
$\Delta E_x \sim 2\,Ze n_{i0} v_{\mathrm{HB}}/(3\,\epsilon_0 \omega_{pi}) \simeq \tfrac{\sqrt{2}}{3}\,a_0 m_e c \omega_0/e$.
Recalling that the (normalized) maximum electrostatic field is $e E_{x,{\rm max}}/(m_e c \omega_0) \simeq \sqrt{2} a_0$,
one then finds that the relative amplitude of the electrostatic field oscillations are of the order 
$\Delta E_x/ E_{x,{\rm max}} \simeq 1/3$. 

This simple estimate turns out to be in very good agreement with our numerical simulations, for example for \reffig{f:piston_osc}(a) we find
$\Delta E_{x,{\rm max}}/E_{x,{\rm max}} \simeq 0.3$.
It is also confirmed by all our simulations performed in the pure HB regime where piston oscillations have been observed,
all of them exhibiting oscillations $\Delta E_x/ E_{x,{\rm max}} \sim 0.3$, independently of the initial plasma density $n_0$
and laser field amplitude $a_0$.

This three-step process suggests that, to set in, piston oscillations require a clear separation between the ion and electron layers,
so that the third stage lasts long enough for the electrostatic field to build up. While this is the case in most of our pure HB simulations,
this clear separation does not hold when considering the dynamic transition regime (for $\tau_r=12\tau_L$). In that case indeed, 
some of the electrons that escape into the vacuum during the initial stage interact with the
standing wave and form energetic bunches through beatwave heating~\cite{ghizzo2007}; they then  
return to the plasma leading to enhanced electron heating [see \reffig{f:spectra_tr}(b),
where electron spectra are plotted at late interaction time $t=200\tl$] by beam-plasma instabilities,
see \reffig{f:piston_osc_PS} and Supplemental Movie~2~\rf{hbsitSuppl}.
This electron heating actually prevents the formation of the double layer with clearly separated ion and electron layers,
as can be seen in  Fig.~\ref{f:piston_osc_PS}(b) for the dynamic transition regime, in contrast with Fig.~\ref{f:piston_osc_PS}(a) for the 
pure HB regime. This henceforth prevents piston oscillations to set in, as is confirmed 
in Fig.~\ref{f:piston_osc}(b,d,f) where none of the three stages discussed for the pure HB case are observed. 
In that case indeed some residual oscillations in the maximum electrostatic field, albeit with a decreased amplitude 
$\Delta E_{x,\mathrm{max}}/E_{x,\mathrm{max}}\simeq0.15$, can be observed. 
Their irregular nature prevents a strong imprint on the ion energy spectrum as they cannot not coherently 
contribute to acceleration or deceleration of 
the fast ions around their mean velocity, see \reffig{f:piston_osc_PS},
and explains the smaller energy spread in the fast ion spectrum observed in Fig.~\ref{f:spectra_tr}(c).

As a result, operating in the \transient\ regime may allow to produce ion beams via HB with a
low energy spread. In contrast to operating in the pure HB regime at lower intensity (or conversely larger density),
a situation which has been shown not to be prone to piston oscillation~\cite{schlegel_POP_2009}, small energy dispersion 
is here obtained without sacrificing mean energy. 

Let us finally note that the effect of electron heating to prevent piston oscillations was also discussed in a previous work~\cite{wu2013}. 
In that case however, the authors relied on the use of elliptically polarized light to allow for electron heating to set in. 

}

}

\begin{figure*}
  \includegraphics[width=0.8\textwidth]{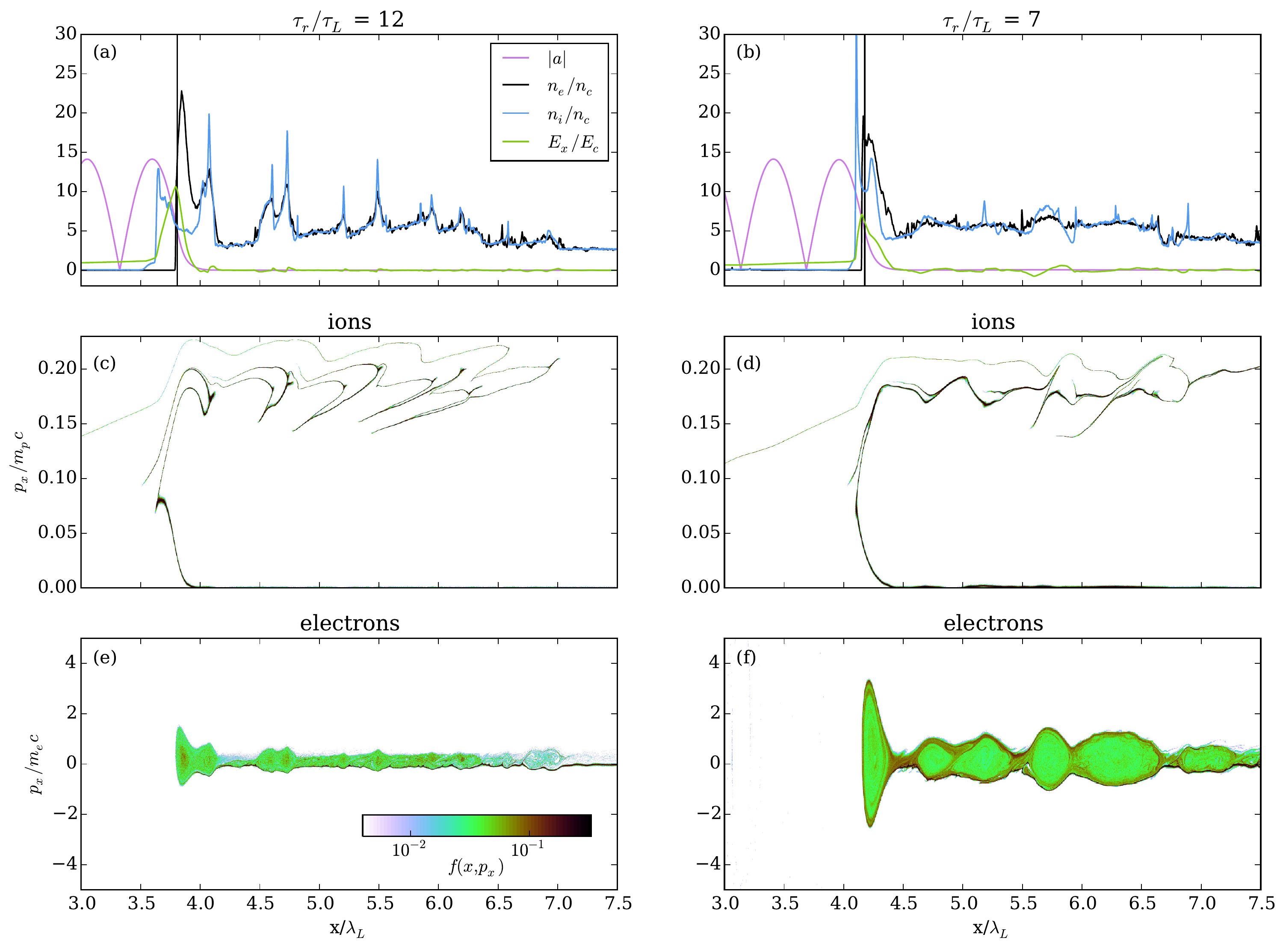}
  \caption{\label{f:piston_osc_PS} Snapshots at $t=45\tl$ of electron $n_e$ and ion density $n_i$, 
  normalized vector potential $|a|$ and longitudinal electric field 
  $E_x$ normalized to the Compton field $E_c = m_e\,c\,\omega_L/e$ for simulations with hydrogen 
  and $a_0=10$, $n_0=2.7\,n_c$ and (a) $\tr=12\tl$ (\hb), (b) $\tr=7\tl$ (\transient\ regime).
  Corresponding ion and electron phase space are shown in (c,d) and (e,f), respectively.
  Movies of the evolution are shown as Supplemental Material~\rf{hbsitSuppl}.
  }
\end{figure*}

\begin{figure*}
  \includegraphics[width=\textwidth]{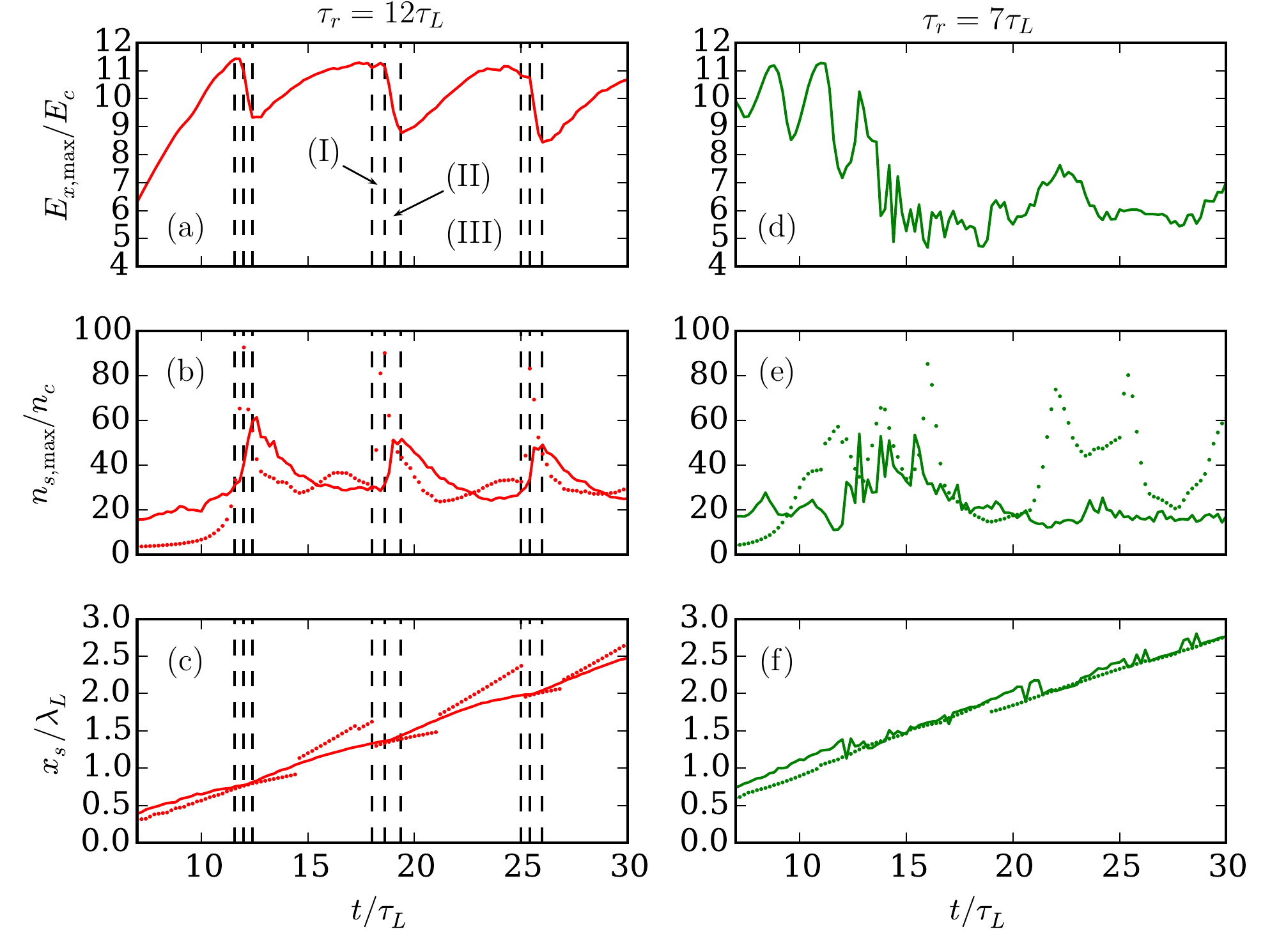}
  \caption{\label{f:piston_osc} Difference in piston oscillations 
  for simulations with hydrogen 
  and $a_0=10$, $n_0=2.7\,n_c$
  for the two values of the rise time $\tr=12\tl$ (\hb, left column) and $\tr=7\tl$ 	(\transient\ regime, right column).
  (a,d) Maximum longitudinal electric field 
  $E_{x,max}$ normalized to the Compton field $E_c = m_e\,c\,\omega_L/e$ versus 	time.
  (b,e) Maximum density $n_{s,max}$ for electrons (solid lines) and ions (dots).
  (c,f) Position of the maximum of the density spike $x_s$ for electrons (solid 	lines) and ions (dots).
  }
\end{figure*}

\section{\label{s:discuss}Discussion and conclusions}

{
Before concluding on this work, we wish to briefly stress that various ion acceleration mechanisms have been identified 
in near-critical plasmas, which are clearly different from the ion acceleration process in the 
dynamic transition regime discussed here. At the boundary of the \rsit\ regime $n_0\simeq n_{th}$
an energetic ion bunch can be formed and 
accelerated to energies much higher than expected from a pure \hb\ scenario,
as discussed in \refref{robinson2011}.
In \refref{weng2012} an incomplete hole-boring 
regime has been reported, which 
occurs for much larger intensities ($a_0\simeq100$) when $v^{\infty}_{\mathrm{SIT}}\simeq\vhb$.
Finally, in the presence of a long enough pre-plasma, trace light ions
can be accelerated by the charge separation field in the pre-plasma\rf{sahai2013,sahai2014}.
In all these regimes ion spectra scale differently than those obtained
in the dynamic transition regime, which follows the usual \hb\ scaling.
}

To conclude, we have studied the transition from the opaque {(HB)} to a transparent {(RSIT)} regime in
the interaction of relativistic laser pulses with plasmas using a combination of PIC
simulations and Hamiltonian dynamics.
The transition to \rsit\
is {found to be} linked to an instability of the plasma-vacuum interface triggered 
by fast electron generation during the early stages of the interaction, 
{as revealed by studying single-electron separatrices}.
Remarkably, this instability can be saturated by an ion-motion-induced deepening of 
the trapping potential at the plasma boundary.
We therefore find that ion motion is involved in a transition 
which is commonly thought of as a purely electron effect.
{As shown in \refsect{s:sep}, this occurs because the strong electrostatic field $E_{x,\mathrm{max}}\propto a_0$ 
at the charge separation layer causes ion response on a time-scale shorter than 
the naive $2\pi\omega_{pi}^{-1}$ estimate. An upper bound for this time-scale which depends on both $\omega_{pi}$ and, importantly, on $a_0$ has been derived.}

We showed that transient effects are important and identified a new dynamic 
transition regime from \rsit\ to \hb.
Surprisingly, the short, transient \rsit\ phase in this regime 
has a long-lasting impact on the properties of the accelerated ions.
\hb\ spectra in near critical plasmas suffer from broadening due to
periodic piston oscillations. {We analyzed these oscillations for non-relativistic 
HB as a three-step process 
and estimated the electric field oscillation amplitude to be approximately $30\%$, independently of $a_0$ and $n_0$, 
in very good agreement with PIC simulations.
Enhanced electron heating in the \transient\ regime prevents this three-step process from setting in, 
therefore ameliorating the effect of the oscillations on the ion spectrum.}
As a result an optimal ion spectrum is obtained both in terms of mean energy and energy spread.

The \transient\ regime is characterized
by complex dynamics and in realistic
scenarios further complicating factors such as
{transverse instabilities} may play a role. 
Recently developed optimization strategies drawing on the field of
complexity science, such as those that rely on genetic algorithms to
control adaptive optics~\cite{he2015}, suggest that there is a
potential to operate laser-driven ion acceleration in the dynamic
transition regime despite the inherently complex dynamics at play.

These results are of fundamental importance for
our understanding of relativistic laser-plasma interaction and for a
wide-range of applications, from particle acceleration to fast
ignition, as they open new paths, for example for the optimization of
laser-driven ion beams.

  %%%%%%%%%%%%%%%%%%%%%%%%%%%%%%%%%%%%%%%%%%%%%%%%%%%%%%%%%%%%%%%%%%%%%%%%%%%%%%%%
  \begin{acknowledgments}
  %%%%%%%%%%%%%%%%%%%%%%%%%%%%%%%%%%%%%%%%%%%%%%%%%%%%%%%%%%%%%%%%%%%%%%%%%%%%%%%%
  The authors are grateful to Andrea Macchi and Tim Dubois for fruitful discussions
  and to Joana Martins for a careful reading of the manuscript. ES and MG also thank
  Theo Schlegel and Vladimir Tikhonchuk for early discussions on HB and piston oscillations.
  ES~thanks Stephan Kuschel for help with the use of the package
  \texttt{postpic}~\cite{postpic}.  This work was supported by the Knut
  and Alice Wallenberg Foundation (\textsc{pliona} project) and the
  European Research Council (ERC-2014-CoG grant 647121). MG and ES
  acknowledge the hospitality of the Max Planck Institute for the Physics
  of Complex Systems where this work was initiated.
  Simulations were performed on resources at Chalmers
  Centre for Computational Science and Engineering (C3SE) provided by
  the Swedish National Infrastructure for Computing (SNIC)
  and on resources of the Max Planck Computing and Data Facility at Garching. 
  EPOCH was developed under UK EPSRC grants EP/G054950/1, EP/G056803/1,
  EP/G055165/1 and EP/M022463/1.
  \end{acknowledgments}
  %%%%%%%%%%%%%%%%%%%%%%%%%%%%%%%%%%%%%%%%%%%%%%%%%%%%%%%%%%%%%%%%%%%%%%%%%%%%%%%%


  \end{document}